\documentclass[11pt]{article}
\usepackage[iso-8859-7]{inputenc}
\usepackage{epsfig}
\usepackage{amsmath}
\newcommand{\newc}{\newcommand}
\newc{\ra}{\rightarrow}
\newc{\lra}{\leftrightarrow}
\newc{\beq}{\begin{equation}}
\newc{\be}{\begin{equation}}
\newc{\eeq}{\end{equation}}
\newc{\ee}{\end{equation}}
\newc{\ba}{\begin{eqnarray}}
\newc{\ea}{\end{eqnarray}}
\newc{\n}{\nu}
\newc{\D}{\Delta}
\def\eps{\epsilon}
\def\la{\lambda}

\newc{\nn}{\nonumber}
\begin{document}
\def\firstpage#1#2#3#4#5#6{
\begin{titlepage}
\nopagebreak
\title{\begin{flushright}
        \vspace*{-0.8in}
\end{flushright}
\vfill {#3}}
\author{\large #4 \\[1.0cm] #5}
\maketitle \vskip -7mm \nopagebreak
\begin{abstract}
{\noindent #6}
\end{abstract}
\vfill
\begin{flushleft}

\end{flushleft}
\thispagestyle{empty}
\end{titlepage}}

\def\simlt{\stackrel{<}{{}_\sim}}
\def\simgt{\stackrel{>}{{}_\sim}}
\date{}
\firstpage{3118}{IC/95/34} {\large\bf   D-brane Inspired Fermion Mass Textures}
 {G. K. Leontaris$^{1,2}$, N.D. Vlachos$^3$}
{\normalsize\sl $^1$Department of Physics, CERN Theory Division, CH-1211, Geneva 23, Switzerland\\
{\normalsize\sl $^2$Theoretical Physics Division, Ioannina
University,
GR-45110 Ioannina, Greece}
\\
{\normalsize\sl $^3$Theoretical Physics Division,  Aristotle University, GR-54006 Thessaloniki, Greece}
 }
{ In this paper, the issues of the quark mass hierarchies and the Cabbibo Kobayashi
Maskawa mixing are analyzed in a class of intersecting D-brane configurations with
Standard Model gauge symmetry. The relevant mass matrices are constructed taking into
account the constraints imposed by extra abelian symmetries and anomaly cancelation
conditions. Possible mass generating mechanisms  including perturbative as well as
non-perturbative effects are discussed and specific patterns of mass textures  are
found characterized by the hierarchies of the scales where the various sources contribute.
It is argued that the Cholesky decomposition of the mass matrices is the most appropriate
way to determine the properties of these fermion mass patterns, while the associated  triangular
mass matrix form provides a unified description of all phenomenologically equivalent symmetric
and non-symmetric mass matrices. An elegant analytic formula is derived for the Cholesky triangular
 form  of the mass matrices where the entries are given as simple functions of the mass eigenstates
 and the diagonalizing transformation entries.  Finally, motivated by the possibility of vanishing
 zero Yukawa mass entries in several D-brane  and F-theory constructions due to the geometry
of the internal space, we analyse in detail all possible texture-zeroes mass matrices within the
proposed new context. These new texture-zeroes are compared to those existing
in the literature while  D-brane inspired cases are worked out in detail.
 }

\vskip 2truecm

\newpage

\section{On the fermion mass problem}

One of the most fascinating challenges in gauge theories of fundamental interactions
today, is the implementation of a natural mechanism providing a satisfactory explanation
of the observed hierarchical fermion mass spectrum and quark mixing. Concentrating
 on the hadronic sector in particular,  we know today experimentally with remarkable accuracy
the quark masses  and the Cabbibo Kobayashi Maskawa (CKM) mixing which arises because the Yukawa matrices are not diagonal in flavor space. This mixing determines
the strengths of the transitions between the various quark flavors explaining the
  observations related to the CP-violation in the neutral Kaon system and other interesting
processes involving quark flavor physics~\footnote{For a recent review see~\cite{Antonelli:2009ws}.}.

Since the birth of modern gauge theories and the establishment of the Standard Model, the problem of mass hierarchy and flavor mixing have been tackled in many ways.
Among the various attempts, abelian and several discrete symmetries~\cite{Froggatt:1978nt}
were often `mobilized' to  discriminate fermion families, supplying thus the theory with more or less
 realistic  textures which reproduce  the observed fermion mass spectrum.

It was subsequently shown that such  $U(1)$  family symmetries arise naturally in the
context of  string models~\footnote{See for example, heterotic and in particular 4d-fermionic
constructions~\cite{Antoniadis:1989zy}-\cite{Faraggi:2006bc}.}.
As a result, a generic characteristic of these constructions is that at the tree-level superpotential
 only one fermion generation (usually the third) of Yukawa couplings is allowed. The remaining two
 fermion families
obtain their masses from higher order non-renormalizable (NR) terms when the various  singlet or any
other Higgs fields appearing in the string spectrum obtain vacuum expectation values (vevs), breaking
thus the surplus $U(1)$ symmetries and filling in the tree-level zeroes of the mass matrices  with
subleading mass terms. As the latter are mainly correlated to the lighter generations a consistent
quark mass hierarchy arises in a natural way.

Searching  for simplicity and maximal predictability on the fermion mass problem, a
 purely phenomenological approach  restricted  to symmetric matrices only, revealed that
 admissible fermion mass textures can be classified to five texture-zero mass matrices which
contain all relevant information  and reproduce the  low energy measurements~\cite{Ramond:1993kv}.
All these textures exhibit a hierarchical structure in the sense that  the magnitudes  of the non-zero
entries coupled to heavier generations are  bigger than those coupled to the lighter ones.

 D-brane models however have paved the way for new  interesting possibilities. Recently, a closer  look at the
 phenomenological properties of the predicted superpotential terms has revealed that completely
 novel structures of non-symmetric mass matrices may appear~\cite{Ibanez:2008my},\cite{Leontaris:2009ci},\cite{Anastasopoulos:2009mr},\cite{Cvetic:2009yh},\cite{Kokorelis:2008ce},\cite{Kiritsis:2009sf},\cite{Cvetic:2009ez}.

In a wide class of these constructions this mainly happens because of constraints
 originating not only from $U(1)$ symmetries, but also from restrictions imposed by  tadpole and
 other anomaly cancelation conditions. For example, in D-brane models built in the context of the
Standard Model symmetry, quark and lepton fields should be distributed between equal numbers of $N$
 and $\bar N$ representations. If we confine ourselves to cases of D-brane configurations
 with the minimal SM spectrum, we find that  the consrtaints are automatically  satisfied for the $SU(3)$
 color group,  however,  the  implementation for the case of the $SU(2)$  doublets
 imposes additional restrictions  on the Yukawa sector. These restrictions lead to rather peculiar mass matrix structures  where the magnitudes of the non-zero entries do not follow a hierarchical
pattern in the sense that was described above.

 Some of these models may prove to be ephemeral, but, they undoubtedly indicate that there are
lots of surprises on the way, thus a detailed analysis towards a complete classification
of mass matrix textures consistent with the fermion masses and mixing is needed.

In the present paper, motivated by recent activity on D-brane phenomenological
explorations related to the origin of fermion masses and their hierarchies,  we
elaborate on the issue of fermion mass spectrum in  SM D-brane variants and develop a
method to construct new viable mass  textures, concentrating mainly on the quark sector.
Our results, are more general and can be applied to the charged lepton and  the neutrino
sector as well. We treat symmetric and non-symmetric mass matrices on an equal footing, by
working out the triangular  (Cholesky) form of the admissible  mass matrices which encodes all
the physical properties in a unique way. It is shown that all matrix elements can be
analytically expressed   in terms of simple unique functions  of the quark masses and the corresponding elements of
 diagonalizing matrix. This latter (Cholesky) form of the $3\times 3$ fermion mass textures can
 be considered to act as a  {\it progenitor} of equivalent classes of admissible symmetric and
 non-symmetric matrices  connected  by  orthogonal  matrices acting on it.

  We further pursue  our  approach by using the Cayley-Hamilton theorem to develop a new more compact formalism
 for the orthogonal transformations that facilitates the analysis of the diagonalizing matrices and
 reveals the geometrical nature of the multiplication properties on
computations regarding  the Cabbibo-Kobayashi-Maskawa  mixing and the quark mass spectrum.
Thus geometrical treatment provides also the tools to investigate cases where up and down quark
matrices are  misaligned and considerable adjustments are necessary to obtain the CKM mixing.
This happens for example in F-theory constructions~\cite{Beasley:2008dc},\cite{Beasley:2008kw}, when
 matter curves for up and down quarks intersect at different points~\cite{Heckman:2008qa}-\cite{Randall:2009dw}.

In the present analysis we will not deal with  corrections attributed to renormalization
group evolution. Thus, for demonstration purposes, experimentally
measured quantities (like masses and mixing) at the electroweak scale will
be used as if the mass matrices were obtained at low energy scales. This is very reasonable
for D-brane models with low unification scale, however, more precise quantitative estimates
for high unification scale scenarios can be easily obtained by taking into account the radiative
corrections which can be easily parametrized in terms of  one parameter only~\cite{Ramond:1993kv}.

 The present paper is organized as follows: In section 2 we derive the Yukawa superpotential
 in  the context of a Standard Model variant emerging from a simple D-brane configuration
 and construct the quark mass matrices with the aforementioned characteristics. A short
 exploration of the magnitude of the Yukawa terms with respect to their particular  origin is
 carried out and a new vector-like parametrization of the matrices is proposed which facilitates
 the subsequent analysis. A characteristic case of the derived quark mass textures is worked
 out is detail and the compatibility of the findings with the low experimental  energy data are
 discussed. In section 3 we introduce the Cholesky form of the mass matrices and explore the
 mathematical properties  which will enable us to classify the admissible quark mass textures.
  We show  how the triangular (Cholesky) form of a mass matric acts as a `progenitor' of
 an equivalent  class of symmetric and non-symmetric mass matrices with the same `physical'
properties, i.e., the same eigenmasses and mixing. Then, in section 4 we introduce a new
parametrization of the orthogonal transformations
and use the analysis of the previous section to express analytically the entries of the triangular
matrix as  functions of the mass eigenstates and the diagonalizing matrix elements. An investigation
on simplified phenomenologically viable texture-zero forms of the triangular matrix is made in
section 5. A separate discussion is also devoted on  comparison issues of the present approach
and the symmetric texture-zeroes in the existing literature. The conditions on the parameter space
in a class of texture-zeroes  matrices to obtain consistency of the D-brane inspired matrices and
examples are worked out in detail in section 6. In section 7 we present our conclusions while in
section 8 we include further details of our calculations.

\newpage

\section{Quark mass textures in a class of intersecting D-brane models}

In order to demonstrate the existence of the novel class of mass matrices in this section we
proceed with the analysis of the Yukawa sector of one particular example based on the simplest
and most economical D-brane configuration which can incorporate the Standard Model (SM) gauge
symmetry. We should point out however, that the peculiar textures derived here
are by no means  a narrow characteristic of  this chosen model. Delving into the variety of the D-brane
SM constructions, one can find  that this specific mass pattern appears in a wide
class of intersecting D-brane SM models~\cite{Ibanez:2008my},\cite{Leontaris:2009ci}, in Gepner
constructions~\cite{Anastasopoulos:2009mr},\cite{Kiritsis:2009sf} as well as in certain GUTs~\cite{Kokorelis:2008ce}.

In all those D-brane analogues of the old successful gauge models,  additional restrictions are imposed
on the matter representations due to the tadpole and anomaly cancelation conditions. More precisely,
for any $U(N_j)$ factor of the gauge symmetry $G_S=\prod_j U(N_j)$, implied by a D-brane configuration,
tadpole cancelation conditions demand equal number of $N_j$ and $\bar N_j$ representations.

In the case of D-brane successors of the Standard Model gauge symmetry~\footnote{Examples of
 D-brane SM analogues with the required restrictions can be found in~\cite{Ibanez:2001nd}.}
with minimal  quark and charged lepton sector, as far as the $U(3)$ representations are concerned,
this requirement is automatically satisfied. Furthermore, in order to implement this condition for the $U(2)$
gauge factor, one has to discriminate between $SU(2)$ doublet and anti-doublet fields and ensure that
 equal numbers are predicted for both  in the massless spectrum. As a consequence, at least  in the simplest
and more appealing cases with the minimal spectrum, not all quark doublet fields arise from the same
intersection, and therefore matter fields belonging to different generations definitely carry
unrelated  $U(1)$ quantum numbers.

In this context, simple hierarchical symmetric textures which were usually discussed in the
literature are far from being realistic and one has to confront the mass texture problem in
a more general context. In this section we demonstrate this fact by giving
one such  simple example implying  representative quark mass textures of these constructions.

\begin{figure}[!t]
\centering
\includegraphics[width=0.28\textwidth]{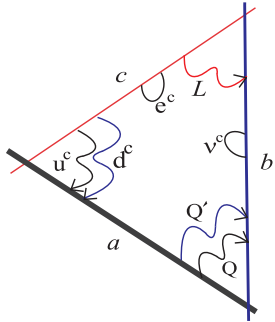}
\caption{\label{fconf}A depiction of the  $U(3)\times U(2)\times U(1)$
D-brane configuration with strings representing the SM states. For the sake of simplicity,
 $a,b,c$ denote  $U(3), U(2)$ and $U(1)$  branes respectively.  In this figure D-branes are not
distinguished from their corresponding mirrors. Thus, the blue string
representing the quark doublet $Q'$ is stretched  between the $D6_a$ and the mirror
 $D6_{b^*}$. Similarly,  one endpoint of the ``$d^c$-string''
 is attached on the mirror $D6_{c^*}$.} \label{f1}
\end{figure}

We assume a D-brane configuration~\cite{Leontaris:2009ci} with three  stacks (call them $a,b,c$)
which generatethe $U(3), U(2),U(1)$, gauge symmetries respectively (the relevant D-brane
configuration  is depicted in figure \ref{f1}).
These are sufficient to incorporate
 the Standard Model gauge symmetry together with its  minimal fermion and Higgs spectrum  which
is shown in Table \ref{U321}.
This consists of the three SM fermion generations, enlarged by the corresponding
right handed neutrinos and one pair of Higgs doublets. It can be checked that anomaly cancelation
conditions are also satisfied.
\begin{table}[!t]
\centering
\renewcommand{\arraystretch}{1.2}
\begin{tabular}{ccrrrr}
\hline
Inters. & $SU(3)\times SU(2)$& ${\cal Q}_a$ & ${\cal Q}_b$ & ${\cal Q}_c$ &$Y$\\
\hline
 $ab$   & $1\times Q\,(3,\bar 2)$ & $1$ & $-1$ & $0$ &$\frac 16$   \\
 $ ab^*$    & $2\times Q'\,(3, 2)$ & $1$ & $1$ & $0$& $\frac 16$   \\
$ac$& $3\times u^c\,(\bar 3,1)$ & $-1$ & $0$        & $1$&$-\frac 23$\\
$ ac^*$ &$3\times d^c\,(\bar 3,1)$ & $-1$ & $0$ & $-1$ &$\frac 13$\\
$bc$& $3\times L\,(1,\bar 2)$ & $0$ & $-1$ & $1$   &$-\frac 12$     \\
 $cc^*$   & $3\times e^c\,(1,1)$ & $0$ & $0$ & $-2$    &$1$    \\
 $bb^*$    & $3\times \nu^c\,(1,1)$ & $0$ & $-2$ & $0$  &$0$      \\
\hline
 $bc^*$   &      $\left.
                  \begin{array}{r}1\times H_d(1,2)\\
                                  1\times {H}_u(1,\bar 2)
                 \end{array}\right.$
          &      $\left.
                 \begin{array}{r}
                 0 \\
                 0
                 \end{array}\right.$
         &       $\left.
                 \begin{array}{r}1\\
                 -1
                 \end{array}\right.$
         &       $\left.\begin{array}{r}
                 1\\
                 -1
                 \end{array}\right.$
         &       $\left.
                 \begin{array}{r}
                 -\frac 12  \\
                  \frac 12
                  \end{array}
                  \right.$\\
\hline
\end{tabular}
 \caption{The quantum numbers of the SM fermions in the $U(3)\times
U(2)\times U(1)$ brane configuration. The last column is the Hypercharge
$Y=\frac 16{\cal Q}_a-\frac 12{\cal Q}_c$ while three  previous
ones refer to the $U(1)$ charges with respect to the $a,b,c$ brane-stacks.\label{U321} }
\end{table}
Let $Q_p'=(3,2), \; p=1,2$ and $Q=(3,\bar 2)$ the three quark doublets and $u_j^c, d_j^c, j=1,2,3$
the right-handed partners. The tree-level quark and lepton Yukawa couplings of this construction are
 \ba
 {\cal W}\supset \lambda^u_{pj}\,Q_p'\,u^c_j\, H_u
 +(\lambda^d_j\,Q\,d^c_j+\lambda^l_{ij}\,L_i\,e^c_j)H_d
\label{SP}
 \ea
  In the first term, the indices $i,j$ run over all three fermion generations, while $p$  takes
 only two values, not yet assigned to particular fermion generations. Thus, only two quark-doublet
  flavors contribute through tree-level perturbative
 Yukawa couplings to the up-quark mass matrix. The reason  is that the additional $U(1)_a$
 charges carried by the various representations do not allow for a coupling involving the
 representation $Q(3,\bar 2)$. For the same reason tree-level mass terms for the two quark doublets
 do not appear in the down quark mass matrix, since  the down  right-handed quarks couple only to the
 remaining quark doublet. As can be inferred, there is a complementary texture zero structure of the up
 and down quark mass matrices at the perturbative level, in the sense that the zero entries of the first
 are non-zero in the second and vice versa. Of course this structure could not be acceptable since
 it would definitely lead to a zero mass for  one up and two down quarks, thus additional contributions
 should be expected from other sources like those discussed above.  Thus, the zero entries are expected to
 be filled in by elements generated by some other mechanism. The main sources are 1) an additional Higgs doublet
 pair,  2) NR-terms obtained when   singlet Higgs fields are introduced in the spectrum or 3) when
 stringy  instanton  effects are taken into account.

If any of the above mechanisms is implemented,    the following  Yukawa terms could
be included to the superpotential
\ba
{\cal W}'&=&{\lambda'}_j^u\,Q\,u_j^c\,H_u'+{\lambda'}_{pj}^d\,Q'_p\,d^c_j\,H_d'
+\cdots
\label{SPprime}
\ea
where the $j$ and $p$ indices in (\ref{SPprime})  span the flavor numbers exactly as in (\ref{SP}).
These new terms are sufficient to provide the missing entries in the quark mass matrices while
dots refer to other possible generated terms (i.e.  Dirac-type neutrino masses etc) which do not
concern us here. The crucial observation however, is that the order of magnitude of these new terms
(\ref{SPprime}) is expected to differ from  those of (\ref{SP}), since their origin is different.

 Taking into account the tree-level perturbative and the additional  terms (\ref{SPprime}),
 the up and down quark mass textures are  classified into three distinct classes depending
 on the particular family assignment. Hence, the first class arises  assuming that $Q'(3,2)$
  accommodates the lightest generation so we have
 \ba
\frac{ m_{Q}}{\langle H_u\rangle}=\left(
\begin{array}{lll}
\kappa_u\eta_{11}^u &\kappa_u\eta_{12}^u & \kappa_u\eta_{13}^u \\
 \la_{21}^u &\la_{22}^u & \la_{23}^u \\
  \la_{31}^u &\la_{32}^u & \la_{33}^u
\end{array}
\right),\;\frac{m_{D}}{\langle H_d\rangle}=\left(
\begin{array}{lll}
  \la_{11}^d &\la_{12}^d & \la_{13}^d\\
  \kappa_d\eta_{21}^d &\kappa_d\eta_{22}^d & \kappa_d\eta_{23}^d\\
\kappa_d\eta_{31}^d &\kappa_d\eta_{32}^d &\kappa_d \eta_{33}^d
\end{array}
\right)\label{c1}
\ea
where, for later convenience, we have written ${\lambda_j'}^u=\kappa^u\eta^u_{1j}$
and ${\lambda'}_{pj}^d=\kappa^d\eta_{pj}^d$ with $p=2,3$ and $j=1,2,3$. The remaining
 two possibilities  are
\ba
\frac{ m_{Q}}{\langle H_u\rangle}=\left(
\begin{array}{lll}
\la_{11}^u &\la_{12}^u & \la_{13}^u \\
\kappa_u\eta_{21}^u&\kappa_u\eta_{22}^u & \kappa_u\eta_{23}^u \\
  \la_{31}^u &\la_{32}^u & \la_{33}^u
\end{array}
\right),\;\frac{m_{D}}{\langle H_d\rangle}=\left(
\begin{array}{lll}
  \kappa_d\eta_{11}^d &\kappa_d\eta_{12}^d & \kappa_d\eta_{13}^d\\
   \la_{21}^d &\la_{22}^d & \la_{23}^d\\
\kappa_d\eta_{31}^d &\kappa_d\eta_{32}^d &\kappa_d \eta_{33}^d
\end{array}
\right)\label{c2}
\ea
and
\ba
\frac{ m_{Q}}{\langle H_u\rangle}=\left(
\begin{array}{lll}
\la_{11}^u &\la_{12}^u & \la_{13}^u \\
\la_{21}^u &\la_{22}^u & \la_{23}^u\\
\kappa_u\eta_{31}^u &\kappa_u\eta_{32}^u & \kappa_u\eta_{33}^u
\end{array}
\right),\;\frac{m_{D}}{\langle H_d\rangle}=\left(
\begin{array}{lll}
  \kappa_d\eta_{11}^d &\kappa_d\eta_{12}^d & \kappa_d\eta_{13}^d\\
 \kappa_d\eta_{21}^d &\kappa_d\eta_{22}^d &\kappa_d\eta_{23}^d\\
   \la_{31}^d &\la_{32}^d & \la_{33}^d
\end{array}
\right)\,.\label{c3}
\ea
Thus, it is clear that all these mechanisms are expected to generate non-symmetric mass matrices with
rather peculiar structure.  In particular the first  additional contributions could arise due
to the second Higgs doublet pair  which can appear in the intersection of branes $D6_b$ and $D6_c$
with the quantum numbers shown in Table~\ref{T2}.
 \begin{table}[!t]
\centering
\renewcommand{\arraystretch}{1.2}
\begin{tabular}{ccrrrr}
\hline
Inters. & $SU(3)\times SU(2)$& ${\cal Q}_a$ & ${\cal Q}_b$ & ${\cal Q}_c$&$Y$ \\
 $bc$   & $\left.\begin{array}{r} H'_d(1,2)\\
                 H'_u(1,\bar 2)
                 \end{array}\right.$  & $\left.\begin{array}{r}0\\
                 0
                 \end{array}\right.$ &$\left.\begin{array}{r}-1\\
                 1
                 \end{array}\right.$& $\left.\begin{array}{r}1\\
                 -1
                 \end{array}\right.$
                 &       $\left.
                 \begin{array}{r}
                 -\frac 12  \\
                  \frac 12
                  \end{array}
                  \right.$ \\
\hline
\end{tabular}
 \caption{The additional Higgs doublets with their quantum numbers. \label{T2} }
\end{table}
These contributions  could lead to smaller, comparable or even larger entries in the mass matrices,
depending of course on the magnitude of the various Higgs vevs. On the contrary, the second and third
sources, namely, the NR or instanton sources  will  fill in the  remaining entries with rather suppressed
contributions. To get a clear insight of the range of the various matrix elements from these latter
sources, we turn our attention to the parameters $\kappa_{u,d}$ and the scale of Yukawa couplings
$\lambda,\eta$.
\begin{table}
\begin{center}
\begin{tabular}{lccc}
  ${\cal W}$-corrections & $\kappa_u$& $\kappa_d$ & Yukawa couplings  \\
   \hline
 $i$) Perturbative :& $\frac{\langle H_u'\rangle}{\langle H_u\rangle}$ &
 $\frac{\langle H_d'\rangle}{\langle H_d\rangle}$ &${\eta_i}^{u,d}\sim \lambda_i^{u,d}$\\
$ii$) Non-renormalizable :& $<1 $& $<1 $&  ${\eta_i}^{u,d}\sim \lambda_i^{u,d}$ \\
$iii$) Non-perturbative :& 1 & 1 &  ${\eta_i}^{u,d}\ll \lambda_i^{u,d}$ \\
\end{tabular}
\end{center}
 \caption{The parameters entering the corrections to the superpotential $i$) from the additional
 Higgs representations and $ii$) from NR-terms and $iii$) from instanton induced terms. \label{T3}}
\end{table}

The non-renormalizable terms in particular are always suppressed by powers of Higgs vevs divided by
some large mass scale $M_S$, being in general of the form
\ba
{\cal W}_{NR}&\supset&\frac{1}{M_S^K}\prod_j^K\langle\Phi_j\rangle\, \eta^u_{ab}\,Q_au_b^c H_u+\frac{1}{M_S^L}\prod_j^L\langle\Phi'_j\rangle\, \eta^d_{ab}\,Q_ad_b^c H_d\nn
\\
&=&\kappa_u\,\eta^u_{ab}\,Q_au_b^c H_u+\kappa_d\,\eta^d_{ab}\,Q_ad_b^c H_d\nn
\ea
Since we expect $\langle\Phi_j\rangle,\langle\Phi'_j\rangle$ to be sufficiently smaller than $M_S$, we
conclude that in general $\kappa_{u,d}< 1$ in this case. The simplest way to realize such terms  in our
particular construction is to
allow the appearance in the spectrum of an additional singlet pair $\Phi_{(0, 2,0)}+\bar\Phi_{(0,-2,0)}$
 which can be represented  by a string stretched in the intersection of $D6_b$ with its mirror brane $D6_{b^*}$.
 Then, the following  fourth order non-renormalizable terms are permitted by the SM gauge and the three
 global $U(1)$ symmetries
\ba
{\cal W}_{NR}&=&\frac{\langle\Phi\rangle}{M_S}\eta^u_jQ\,u_j^c\,H_u+\frac{\langle\bar\Phi\rangle}{M_S}
\eta^d_{pj}\,Q'_p\,d^c_j\,H_d\label{newY}
\ea
which contribute exactly to the zero entries of the tree-level mass matrices discussed above.

Finally, we discuss in brief the non-perturbative contributions. It has been
suggested~\cite{Blumenhagen:2006xt,Ibanez:2006da} that in intersecting D-brane models, several
missing tree-level Yukawa couplings could be generated from non-perturbative effects.
In the present model in particular, considering ${\cal E}2$ instantons in type IIA string theory having appropriate
number of intersections with the $D6$-branes,  non-perturbative terms of the form~\cite{Leontaris:2009ci}
\ba
{\cal W}_{n.p.}&\propto&\,e^{-S_{\cal E}}Q\,u_j^c\,H_u+e^{-S_{{\cal E}'}}\,Q'_p\,d^c_j\,H_d\label{instcoup}
\ea
are induced, where the instanton action $S_{\cal E}$ can absorb the $U(1)$ charge excess of the matter
fields operator involved, so that the whole coupling is totally gauge invariant.
The induced couplings (\ref{instcoup}) involve an exponential suppression by the classical instanton action
${\cal W}_{n.p.}\propto \exp\{-\frac{8\pi^2{\rm Vol}_{\cal E}}{g_a^2{\rm Vol}_{D6_a}}\}$ which, as can  be
seen,  depends on the volume Vol${}_{{\cal E}}$ of the cycle wrapped by the instanton and is inversely
proportional to the perturbative string gauge coupling $g_a^2$. Thus, as opposed to the tree-level
$\lambda$-Yukawa couplings, the $\eta$ couplings in the non-perturbative case exhibit  a  significant
suppression, therefore the corresponding lines of the matrices are substantially suppressed with  respect
to the tree-level contributions  ($\lambda$ couplings). We say in this case that the up  and down quark mass
matrices exhibit a complementary structure in the sense that the small elements  in the first matrix occupy
the entries where there are large ones in the second matrix and vice versa.

Let us finally point out that in the perturbative case, we have a variety of possibilities, depending on the ratios
 of the Higgs vevs  $\kappa_{u,d}$. We particularly mention the interesting case   $\kappa_{u}>1$ and  $\kappa_{d}<1$ where the up and down quark mass matrices are `aligned' in the sense that both of them exhibit the same
hierarchical structure. For example, in the first set given in (\ref{c1}),  large elements occupy the  first
matrix line of both up and down quarks, while smaller entries are in the two remaining lines.  For
$\kappa_{u}<1$ and  $\kappa_{d}>1$ the opposite is true.
 The above remarks apply also analogously for the remaining two classes of textures presented above.

\subsection{On the structure of the D-brane inspired mass matrices}

Motivated by the observations discussed in  detail above, in the present section we introduce a
new formalism which is suitably adapted to the D-brane inspired fermion mass textures  discussed
in the previous section. Indeed, observing the structure of the mass matrices (\ref{c1}-\ref{c3}),
 we deduce that  the relative order of magnitude of elements of different matrix-lines are determined
 by the particular mechanism employed. Some of the entries of course  could be accidentally zero due
 to some kind of symmetry, hence leading to some kind of non-symmetric texture-zero cases, but even so, the
 scale of the generating mechanism  is set by the remaining non-zero elements of the particular line.
Therefore, we find that the analysis of the  $3\times 3$ fermion mass  matrices of this specific kind is largely
 facilitated by treating the elements of each matrix line as a three-component vector. In this case, it
 is the magnitude of the vector rather than the individual couplings themselves that should be correlated
  to the specific source  these couplings  arise.

To set up our formulation and make our analysis as clear as possible, we start with a specific case,
namely the up and down quark mass textures (\ref{c1}) of the previous section.
Consider thus the down quark mass texture where tree-level perturbative contributions are assumed
for the elements $m_{D_{11}}, m_{D_{12}}, m_{D_{13}}$, while instanton induced or NR subleading
terms or second Higgs contributions fill in the rest of the mass matrix. We distinguish the
two kinds of contributions  with Latin and Greek letters respectively~\footnote{For convenience
we simplify the notation $m_{D_{1j}}\ra x_{1j}$ and so on. We further restrict our analysis in real
mass matrices. This does not affect our main conclusions while the generalization to complex matrices
is straightforward. }
\ba
m_D&=&\left(
\begin{array}{lll}
 x_{11} & x_{12} & x_{13} \\
 \zeta _{21} & \zeta _{22} & \zeta _{23} \\
 \kappa _{31} & \kappa _{32} & \kappa _{33}
\end{array}
\right)
\ea
We represent the line elements as the following 3-component vectors,
\ba
\vec x&=&(x_{11},x_{12},x_{13})\nn\\
\vec \zeta&=&(\zeta _{21} , \zeta _{22} ,\zeta _{23})\nn\\
\vec \kappa &=&(\kappa _{31} , \kappa _{32} ,\kappa _{33})\nn
\ea
According to our discussion, the magnitudes $|\vec \zeta|$ and $|\vec \kappa|$  are
 expected to be defined at some common scale which in general differs from that of $|\vec x|$.
 Since the matrices are non-symmetric, for the diagonalization procedure
 we form the  symmetric  matrix  $M_{D}^2=m_Dm_D^T$  which in vector like form is written
\ba
M_{D}^2=m_Dm_D^T  &=&\left(
\begin{array}{lll}
 \vec x\cdot\vec x &  \vec x\cdot\vec\zeta & \vec x\cdot\vec\kappa \\
  \vec x\cdot\vec\zeta &\vec \zeta   \cdot\vec\zeta & \vec \zeta   \cdot\vec\kappa  \\
 \vec x  \cdot\vec\kappa &  \vec \zeta   \cdot\vec\kappa &  \vec \kappa   \cdot\vec\kappa
\end{array}
\right)\label{vecdown}
\ea
Thus, in case $\vec x$ and $\vec \zeta,\vec \kappa$ define two substantially different scales, the entries
of the matrix (\ref{vecdown}) in general could belong to three categories:
In the case of instanton corrections for example, we expect that $|\vec x|\gg |\vec\zeta|,|\vec\kappa|$,
so that the biggest element is ${M_D^2}_{11}\equiv |\vec x|^2$, with the other two diagonal
 entries ${M_D^2}_{22}=|\vec\zeta|^2>0$ and ${M_D^2}_{33}=|\vec\kappa|^2>0$ being substantially smaller.
The off-diagonal elements determined by the inner products $\vec x\cdot\vec\zeta, \vec x\cdot\vec\kappa$,
are expected to be at most at some intermediate scale, while $\vec\zeta\cdot\vec\kappa$ might be even smaller.
It is possible of course that some inner products are zero, i.e., $\vec \zeta   \cdot\vec\kappa=0$ etc
which essentially implies that the two vectors are orthogonal. Certainly, as we have already discussed
in the previous section, the results are similar for the case of NR-contributions however the prospects are
completely different if the entries $\vec \zeta,\vec \kappa$ originate from a second Higgs doublet.
We will comment about this possibility in subsequent sections.

In a similar manner, we may write the up quark matrix
\ba
m_U&=&\left(
\begin{array}{ccc}
 \xi _{11} & \xi _{12} & \xi _{13} \\
 y_{21} & y_{22} & y_{23} \\
 z_{31} & z_{32} & z_{33}
\end{array}
\right)
\ea
 and since the matrix is also  non-symmetric, we construct the matrix $m_Um_U^T$
\ba
M_{U}^2=m_Um_U^T  &=&\left(
\begin{array}{lll}
 \vec \xi\cdot\vec \xi &  \vec \xi\cdot\vec y & \vec \xi\cdot\vec z \\
  \vec \xi\cdot\vec y &\vec y  \cdot\vec y & \vec y  \cdot\vec z  \\
 \vec \xi  \cdot\vec z &  \vec y \cdot\vec z &  \vec z \cdot\vec z
\end{array}
\right)\label{vecup}
\ea
with $\vec\xi=(\xi_1,\xi_2,\xi_3)$ and so on. If again we assume that corrections
are small compared to perturbative tree-level terms, then we expect $|\vec x|,|\vec y|\gg |\vec\xi|$
while the magnitudes of the inner products, following analogous reasoning with that of the down quark
mass matrix discussion above, are anticipated at smaller scales. It is worth observing that
the two scales of the non-symmetric mass $m_U$ matrix yield three of them in the symmetric product
$m_Um_U^T$, to which the three flavor-hierarchy can be naturally attributed.
A more involved analysis should be carried out if other sources are included.

 \subsection{ D-brane inspired textures: A case study}
In the previous sections we have seen that D-brane scenarios induce  a variety of
fermion mass  textures where  the hierarchies of their entries depend on the particular
mechanism employed. Of course, not all of these
textures can be compatible with the know data. In the present section, we elaborate
on the consistency of a specific pair of them with the measured low energy data, while
in the next sections we shall develop a more general and novel formalism.

We  start the analysis with the down quark mass matrix (\ref{vecdown}) and the
assumption that only one Higgs pair is included in the spectrum. In this case, there
are only instanton or NR-contributions to the tree-level zero entries, thus we expect
that $\vec x\cdot\vec x\gg \vec x\cdot\vec\zeta, \vec \zeta\cdot\vec\zeta,\vec \kappa
\cdot\vec\kappa$. For  later convenience we define
\ba
r\,\cos\theta&=&-\frac{ \vec x\cdot\vec\zeta}{ \vec x\cdot\vec x}\nn
\\
r\,\sin\theta&=&\frac{ \vec x\cdot\vec\kappa}{ \vec x\cdot\vec x}\nn
\ea
Making use of the fact that the orthogonal transformation does not alter the
physical quantities, we can  arrange that the two vectors $\vec \zeta$ and $\vec\kappa$ are orthogonal, $\vec \zeta\cdot\vec\kappa=0 $,
while to keep the algebra tractable, without loss of generality we assume a slightly simplified  texture and at
first approximation we may put their magnitudes equal
$\vec \zeta\cdot\vec\zeta=\vec \kappa\cdot\vec\kappa= s^2$ to obtain
\ba
m_{D}m_{D}^T&=&\left(
\begin{array}{ccc}
 1 & -r \cos (\theta ) & r \sin (\theta ) \\
 -r \cos (\theta ) & s^2 & 0 \\
 r \sin (\theta ) & 0 & s^2
\end{array}
\right)m_0^2\label{downpar}
\ea
This has to be  diagonalized by an orthogonal matrix $V_d$, to
give a diagonal matrix with elements the mass eigenstates squared
\ba
V_d^T(m_{D}m_{D}^T)V_d&=&\left(M_D^{2}\right)_{diag.}=\left(
\begin{array}{lll}
 m_d^2 & 0 & 0 \\
 0 & m_s^2 & 0 \\
 0 & 0 & m_b^2
\end{array}
\right)\label{mmT}
\ea
Taking into account the quark mass hierarchies,  the diagonalizing matrix can be approximated by
\ba
V_d &\approx&\left(
\begin{array}{lll}
 -\frac{m_s}{m_b} &0 &1-\frac{m_s^2}{2 m_b^2}
   \\
 -\left(1-\frac{m_s^2}{2 m_b^2}\right)  \cos (\theta ) & \sin (\theta ) & -\frac{m_s}{m_b}\cos (\theta )  \\
 \left(1-\frac{m_s^2}{2 m_b^2}\right)\sin (\theta ) &\cos (\theta )  & \frac{ m_s}{m_b}\sin (\theta )
\end{array}
\right)\nn
\ea
We can write a convenient approximate form for the preceding down quark matrix
\ba
m_{D}m_{D}^T&\sim&\left(
\begin{array}{lll}
 1& -\xi\cos\theta & \xi\sin\theta \\
-\xi\cos\theta & \xi^2 & 0 \\
 \xi\sin\theta & 0 & \xi^2
\end{array}
\right)m_b^2\nn
\ea
where $\xi$ is a known function of down quark mass ratios
($\xi\sim 3.1\times 10^{-2}$) given in the appendix.

Up to this point, we have expressed analytically   the down quark entries of the first texture (\ref{c1})
as functions of the mass eigenstates (down quark mass ratios) and the mixing.
Next, we can use this result and the known CKM matrix to derive the admissible up
quark mass matrix which should be compared with the findings of section 2.
We may facilitate the analysis by using the Wolfenstein
parametrization~\cite{Wolfenstein:1983yz} for the CKM matrix (see appendix for conventions).
Thus, having determined $V_d$ while using the relation $V_u=V_dV_{CKM}^\dagger$
we  construct first the diagonalizing matrix $V_u$ of the up quarks. Assigning
$\left(m_U^{2}\right)_{diag.}$ the diagonal matrix with diagonal elements $\{m_u^2,m_c^2,m_t^2\}$,
we can use the relation
\ba
{\cal M}_{U}^2\equiv m_Um_U^{\dagger}&=&V_u^{\dagger}\left(m_U^{2}\right)_{diag.}V_u
\ea
to determine analytically all up-quark entries.  Putting
\ba
\epsilon&=&A \sin (\theta ) \lambda ^2+\xi  \cos (\theta )\nn\\
\epsilon'&=&\xi  \sin (\theta )-A \lambda ^2 \cos (\theta )\nn
\ea
where $\lambda\approx 0.2357$ and $A\sim {\cal O}(1)$ the well known parameters of the Wolfenstein
parametrization of CKM, we obtain
\ba
m_{U}m_{U}^T&\approx&\left(
\begin{array}{lll}
 1& -\eps  & \eps' \\
 -\eps & \eps^2 & \eps\eps' \\
\eps'   & \eps\eps'  & {\eps'}^2
\end{array}
\right)\,m_t^2\label{alignedup0}
\ea
Since the parameters in (\ref{alignedup0}) are  $\eps,\eps'<1$, 
we infer that the resulting up-quark mass matrix structure is compatible with
the aligned scenario discussed in the end of the previous section. As already explained, this alignment
can for example occur in the presence of a second Higgs pair. In other words, the present analysis shows
that in simple D-brane Standard Model scenarios with minimal spectra as is the  case under consideration,
instanton effects are not enough to reproduce the known quark mass hierarchies in both, down and up quark
sectors. In this particular example we have worked out it was shown that the very precise form of
the CKM matrix requires also  the up-quark mass matrix to be aligned with that of the down quarks and this can
happen if additional contributions of a second Higgs doublet are included. Of course, this specific
example does not exhaust all the possibilities. In constructing D-brane models with more complicated symmetries
and matter spectra, rather involved Yukawa textures appear, therefore, in the subsequent, we explore
systematically  general non-symmetric mass matrices and classify the admissible cases.

\section{The Cholesky form of the Mass Matrix}

The classification of all (symmetric and non-symmetric) admissible $3\times 3$ mass matrices
which reconcile the known quark mass hierarchy and mixing is a rather hard task. For example,
the  symmetric squared  matrix $m_Dm_D^T$ discussed above could emerge from a variety
of symmetric or non-symmetric $m_D$ textures. To pursue further this issue and find the
admissible $m_D$ textures, we first rely  on the observation that all eigenvalues of the mass matrix
squared are positive and the fact that any positive definite symmetric matrix can be decomposed
into a product of a lower triangular matrix times its conjugate transpose. The lower triangular
matrix can be identified with the mass matrix $m_D$ or $m_U$ respectively (Cholesky decomposition).
 We will show that this triangular matrix contains all the necessary information related to the fermion mass
 eigenstates and mixing angles of a whole class of matrices. Indeed, once the triangular matrix is specified,
 an `equivalent' class ${\cal C}$ of matrices with the same `physical' properties can be generated when
we multiply the latter by  an orthogonal  matrix. More precisely,  the eigenmasses and
 the eigenvectors of corresponding $mm^T$  of matrices $m\in {\cal C}$ are the same.  We call the
 triangular form of the $3\times 3$ mass matrix a  {\it progenitor}.

On our general exploration for  the admissible quark mass textures, we start this section with
the derivation  of some mathematical formulae relating the mass matrices to their Cholesky progenitor
and the corresponding orthogonal transformation, which are going to be useful in the subsequent analysis.
From (\ref{mmT}) we have seen that the general symmetric mass matrices to be diagonalized are
of the form
\begin{equation}
mm^{T}=UM_{diag.}^{2}U^{T}
\end{equation}%
where the $m$ matrix stands for the up or down quark case, while $U$ is the corresponding  orthogonal transformation and $M_{diag}^2$ is the corresponding diagonalized  (up or down) quark matrix squared
\begin{equation}
M_{diag.}^{2}=%
\begin{bmatrix}
m_{1}^{2} & 0 & 0 \\
0 & m_{2}^{2} & 0 \\
0 & 0 & m_{3}^{2}%
\end{bmatrix}%
~\cdot
\end{equation}%
Since $mm^{T}$ is positive definite and symmetric there exists a Cholesky decomposition
\begin{equation}
M_CM_C^{T}=mm^{T}\label{CF}
\end{equation}%
where the $Cholesky $ lower triangular form is written
\begin{equation}
M_C=%
\begin{bmatrix}
a_1 & 0 & 0 \\
b_1 & b_2 & 0 \\
c_1 & c_2 & c_3%
\end{bmatrix}%
\label{Chol}
\end{equation}%
From (\ref{CF}) we have
\ba
m^{-1}M_CM_C^{T}\left( m^{T}\right) ^{-1}=\;I\;=m^{-1}M_C\left( m^{-1}M_C\right) ^{T}
\label{CO}
\ea
where $I$ stands for the $3\times 3$ unit matrix. From the last equality
we deduce that the matrix $m^{-1}M_C$  is equivalent to an orthogonal matrix $U_{M}~$, i.e.,
the original matrix is connected to its Cholesky form by the relation
\ba
m=M_CU_{M}\label{mSy}
\ea
 Note that to any equivalent class of matrices, there also exists an
 associated symmetric
matrix $M_{s}~$(not uniquely defined) which satisfies the relation
\ba
M_{s}M_{s}=mm^{T}\nn
\ea
Since the latter can also be written as
\begin{equation}
m^{-1}M_{s}M_{s}\left( m^{T}\right) ^{-1}=m^{-1}M_{s}\left(
m^{-1}M_{s}\right) ^{T}=I\nn
\end{equation}%
we conclude that $m^{-1}M_{s}$ is also an orthogonal matrix, i.e.,
the associated symmetric mass matrix is connected to the Cholesky form
\ba
m=M_{s}U_{s}^{T}~\label{mSyU}
\ea
where $U_{s}$ is also orthogonal. Thus, equating (\ref{mSy}) and (\ref{mSyU}) we get
\begin{equation}
m=M_CU_{M}=M_{s}U_{s}^{T}
\end{equation}%
These relations allow us to connect the symmetric matrix to the original one by
\begin{equation}
M_{s}=M_CU_{M}U_{s}~\cdot
\end{equation}%
We see that all mass textures (symmetric and non-symmetric) having the same
physical properties (mass eigenvalues and mixing) can be constructed
by multiplying the Cholesky matrix with an orthogonal matrix. The corresponding
symmetric matrix $M_{s}$ can be easily constructed from the relation
\begin{equation}
M_{s}=UM_{diag.}U^{T}
\end{equation}%
(it is in fact the square root of $mm^{T}$) where
\begin{equation}
M_{diag.}=\sqrt{M_{diag.}^{2}}=%
\begin{bmatrix}
\pm m_{1} & 0 & 0 \\
0 & \pm m_{2} & 0 \\
0 & 0 & \pm m_{3}%
\end{bmatrix}%
~\cdot
\end{equation}%
Note also that
\begin{equation}
M_C=UM_{diag.}\left( U_{M}U_{s}U\right)^{T}
\end{equation}%
i.e. a biorthogonal transformation diagonalizes $M_C~$. 

Thus, we conclude that
the problem is essentially the factorization  of the square matrix to a lower triangular
form times an orthogonal matrix,  which has a unique solution if the elements of the main diagonal
of $M_C$ are taken to be positive.

\begin{figure}[h]
\centering
\includegraphics[scale=.8]{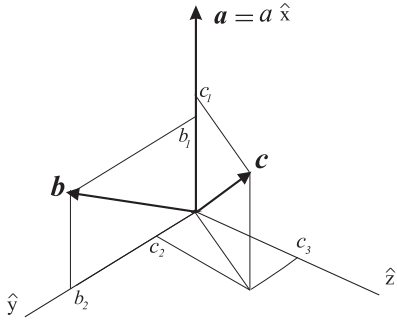}
\caption{The vector representation of the matrix elements. The three lines in the Cholesky
decomposition can be represented by three vectors: The coordinate system
is chosen so that the vector $\vec a=a\hat x$  lies
on the $x$-axis, while $\vec b$ lies on the $(x,y)$-plane. }
 \label{axb1}
\end{figure}

In the subsequent, we will be concerned mainly with real mass matrices, thus all
diagonalizing matrices will be represented  by orthogonal transformations.
The generalization to  complex mass matrices  can be easily done while our main
findings and conclusions do not change. Notice also that in this case
the Cholesky form of the $3\times 3$  matrix   can be conveniently
 visualized (see figure \ref{axb1}) in terms of the vector
representation of the matrices  introduced in the previous section.

Summarizing, we emphasize that once the physical  properties of the triangular matrix have
been explored, any other possible physically equivalent form (symmetric or
non-symmetric texture)  can be obtained by means of an orthogonal transformation which in
standard parametrization reads
\ba
U=\left(
\begin{array}{ccc}
 \cos \alpha  \cos \gamma  & \cos \gamma  \sin \alpha & \sin \gamma  \\
 -\cos \beta  \sin \alpha -\cos \alpha  \sin \beta  \sin \gamma  & \cos \alpha  \cos
   \beta -\sin \alpha  \sin \beta  \sin \gamma  & \cos \gamma  \sin \beta  \\
 \sin \alpha  \sin \beta -\cos \alpha  \cos \beta  \sin \gamma  & -\cos \alpha  \sin
   \beta -\cos \beta  \sin \alpha  \sin \gamma  & \cos \beta  \cos \gamma
\end{array}
\right)\label{Ortho}
\ea
Indeed, since $\left\{M_C{\,U}\right\}\left\{{U\,}^TM_C^T\right\}=M_CM_C^T$,
we can generate equivalent forms by means of the transformation
$ M_x=M_C{\,U}$ as already was proven above. Thus, the first line of $M_x$ for example becomes
\ba
{M_x}_1&=&\{a_1 \cos \alpha  \cos \gamma, a_1 \sin \alpha  \cos \gamma, a_1 \sin \gamma \}\nn
\ea
while analogous, although more lengthy expressions hold for the other two lines too. We therefore
see that the orthogonal matrix rearranges the elements within a given line of the mass matrix,
 but never  mixes them with the elements of the other lines.  Generally, if ${M_x}_i$ is the $i$-th line
 (or a three-component vector in our notation), then the inner product is ${M_x}_i{M_x}_i$
while the orthogonal transformation does not alter the magnitude of the vector, so
\ba
{M_x}_1\cdot {M_x}_1&=&|\vec a|^2=a_1^2\nn\\
{M_x}_2\cdot {M_x}_2&=&|\vec b|^2=b_1^2+b_2^2\label{magn}\\
{M_x}_3\cdot {M_x}_3&=&|\vec c|^2=c_1^2+c_2^2+c_3^2\nn
\ea
Notice also that these quantities coincide with the diagonal elements of $M_CM_C^T$, which
implies that
$$|\vec a|^2+ |\vec b|^2+ |\vec c|^2=m_1^2+m_2^2+m_3^2$$
Also, the product of the entries on the diagonal equals the product of the eigenmasses
$$a_1b_2c_3=m_1m_2m_3$$

For our purposes, the crucial observation here is that the magnitudes of the three vectors
$\vec a,\vec b,\vec c$ which represent the three lines of the matrix, remained unaltered
under the multiplication of $M_C$ by orthogonal  transformations. Therefore,
since any matrix can be cast in triangular form by an orthogonal transformation, we
can concentrate our analysis
in the latter. The peculiar form of the matrices derived in the previous section,
where each line of mass entries is related to a different
mass generating mechanism,  justifies an analogous treatment.

Following the discussion above we attempt now to construct analytically $M_C$
from a general mass matrix and correlate its elements  to those of the  corresponding
triangular form. This procedure  essentially corresponds to the decomposition of a square
matrix to its triangular form times an orthogonal matrix (QL decomposition).
 To this end, we consider the general $3\times 3$ mass  matrix
\ba
m&=&\left(
\begin{array}{ccc}
 m_{11} &  m_{12}  &  m_{13}  \\
  m_{21} &  m_{22}  &  m_{23} \\
 m_{31} &  m_{32}  &  m_{33}
\end{array}
\right)\label{Min}
\ea
Next we define the vectors $\vec\xi_j$, $j=1,2,3$
\ba
\vec{\xi}_j&=& (m_{j1} ,  m_{j2}, m_{j3})
\ea
hence, the matrix can be written
\ba
m&=&\left(
\begin{array}{c}
 \vec{\xi}_1  \\
\vec{\xi}_2 \\
\vec{\xi}_3
\end{array}
\right)\label{Mvec}
\ea

To construct the Cholesky matrix we rely on the Gram-Schmidt orthogonalization procedure and
introduce the following  orthogonal set
\ba
\vec u_i&=&\vec\xi_i-\sum_{j=1}^{i-1}(\vec\xi_i\cdot\hat e_j)\,\hat e_j
,\;i=1,2,3\label{orthov}
\ea
where we have defined the following three orthogonal unit vectors
\ba
\hat e_i&=&\frac{\vec u_i}{|\vec u_i|},\;i=1,2,3\label{univec}
\ea
Using the above formulae we can decompose the original matrix as follows
\ba
\left(
\begin{array}{c}
 \vec{\xi}_1  \\
\vec{\xi}_2 \\
\vec{\xi}_3
\end{array}
\right)&=&\left(
\begin{array}{ccc}
\vec\xi_1\cdot\hat e_1  &0&0\\
\vec\xi_2\cdot\hat e_1  &\vec\xi_2\cdot\hat e_2  &0 \\
\vec\xi_3\cdot\hat e_1  &\vec\xi_3\cdot\hat e_2  &\vec\xi_3\cdot\hat e_3
\end{array}
\right)\left(
\begin{array}{c}
 \hat{e}_1  \\
\hat{e}_2 \\
\hat{e}_3
\end{array}
\right)\label{Chol}
\ea
where the last column defines a orthogonal matrix
\ba
U&=&\left(
\begin{array}{c}
 \hat{e}_1  \\
\hat{e}_2 \\
\hat{e}_3
\end{array}
\right)\label{Ug}
\ea
 whose elements are determined by (\ref{orthov}) and (\ref{univec}). The
 formula (\ref{Chol}) is an explicit realization of (\ref{mSy}) with $U$ given by
 (\ref{Ug}). Using this
 latter representation of the triangular matrix, in the
 next section we will  give  simple analytic formulae of its non-zero entries
 as functions of the mass eigenstates and the diagonalizing matrix elements.

 \section{The rotating matrices}

 We have seen that the triangular form of the mass matrix  plays the r\^ole
 of the progenitor of all classes of symmetric as well as non-symmetric
 mass matrices with the same physical properties. Once the triangular form of the mass matrix is determined,
 its multiplication with an orthogonal mass matrix $U(\alpha,\beta,\gamma)$ leads to
 an equivalent texture which, depending on the  specific choice of the angles
 $\alpha,\beta,\gamma$ of the $U$ matrix may be symmetric or non-symmetric or might have some
 simplified texture-zero form. All matrices obtained through this procedure from
 the same progenitor result to  the symmetric product $mm^T$, thus they are
 characterized by the same diagonalizing orthogonal trasformation and the same eigevalues.

Therefore, the triangular matrix contains all the necessary  information for
the mixing and mass eigenstates. Its usefulness will be evinced  by the fact
that all its entries can be expressed as simple functions of the eigenmasses
and the elements of the diagonalizing matrix. In the basis where one of the quark matrices
is diagonal, all the elements of the other quark sector in its triangular form
are uniquely determined in terms of the mass eigenstates and the CKM mixing.
When up and down quark matrices appear in non-diagonal form, the elements of the
triangular matrices are not uniquely determined. The entries of each matrix can
be expressed in terms of the eigenmasses and the angles of the diagonalizing
orthogonal matrix, whilst, only the elements of the product $V_u^{\dagger}V_d$ of the two
 transformations is constrained to be the CKM matrix. Therefore, as expected, a variety
of mass textures can result to the known CKM mixing.  The triangular form of the mass
textures gives the advantage of treating in a simple and economical manner the
problem of finding the admissible classes of quark matrices which reconcile the
experimental data of mass and CKM mixing.

 In this section we develop a general formalism to  determine all possible
 classes of admissible mass textures  for the  up and down quark sectors.
 As pointed out, our procedure is general and applies equally well in
 the case of the lepton sector.
 We introduce here a novel parametrization of the orthogonal matrices
 and the CKM matrix which  will facilitate the subsequent analysis.
 To this end, let us define the antisymmetric $3\times 3$ matrices
\ba
s_{1}=%
\begin{bmatrix}
0 & 0 & 0 \\
0 & 0 & 1 \\
0 & -1 & 0%
\end{bmatrix},%
\;\;
s_{2}=%
\begin{bmatrix}
0 & 0 & 1 \\
0 & 0 & 0 \\
-1 & 0 & 0%
\end{bmatrix},
\;\;
s_{3}=%
\begin{bmatrix}
0 & 1 & 0 \\
-1 & 0 & 0 \\
0 & 0 & 0%
\end{bmatrix}\label{sigmas}
\cdot
\ea
and a unit vector
\ba
\hat n&=&(n_1,n_2,n_3)
\ea
 Using the Cayley-Hamilton theorem~\cite{Gantmacher},
we find that the exponential of the $3\times 3$ matrices proportional to the inner product
$\hat n\cdot\vec s$ is written
\begin{equation}
\exp \left[ \alpha\widehat{n}\overrightarrow{s}\right] =1+\sin \alpha~\widehat{n}%
\cdot\overrightarrow{s}+\left( 1-\cos \alpha\right) \left( \widehat{n}\cdot\overrightarrow{s%
}\right) ^{2}
\end{equation}%
or equivalently
\begin{equation}
\exp \left[ \alpha\widehat{n}\overrightarrow{s}\right] =1+\sin \alpha~\widehat{n}%
\cdot\overrightarrow{s}+2\sin ^{2}\left( \frac{\alpha}{2}\right) \left( \widehat{n}%
\cdot\overrightarrow{s}\right) ^{2}\cdot
\end{equation}
The general orthogonal matrix can therefore be written as
\begin{equation}
U=\exp \left[ \alpha\widehat{n}\cdot\overrightarrow{s}\right] \label{unit}
\end{equation}%
Indeed, since $\widehat{n}\cdot\overrightarrow{s}$ is traceless, the determinant of $U$ is
unity  and since $\widehat{n}\cdot\overrightarrow{s}$ is antisymmetric it ensures that
$U^{T}U=1$.  Therefore, the orthogonal matrix can be parametrized by the angle $\alpha$ and the
unit vector $\hat n$ whose components constitute the directional cosines along the `directions'
$s_{1,2,3}$.

To get a feeling of this new parametrization  in terms of the `orthogonal'  basis $s_1,s_2,s_3$
 of the orthogonal matrix and its correlation with the standard parametrization, let us consider
 the particular case where the  vector $\hat n$ is aligned along a specific `axis', i.e. let
 $\hat n=(0,0,1)$. Then
 \ba
U(\alpha)&=&\exp \left[ \alpha \,{s_3}\right]\equiv\left(
\begin{array}{lll}
 \cos (\alpha ) & \sin (\alpha ) & 0 \\
 -\sin (\alpha ) & \cos (\alpha ) & 0 \\
 0 & 0 & 1
\end{array}
\right)\nn
 \ea
 Analogous expressions hold for alignments to the remaining two axes.

  For our subsequent
 analysis it is also useful to  express  the CKM matrix  in the basis $s_1,s_2,s_3$.
Plugging in the numerical values of its entries as they are measured by the experiment
(see appendix) we find that the CKM matrix can be written
\begin{equation}
 V_{CKM}=\exp[\phi_c\widehat{n}\overrightarrow{s}]
\end{equation}%
where the angle $\phi_c=0.231505$ and the unit vector components along the
orthogonal  directions $s_1,s_2,s_3$ are
\begin{equation}
\widehat{n}=%
\begin{bmatrix}
0.179694 \\
0.0177102 \\
0.983564%
\end{bmatrix}\label{nCKM}
\end{equation}%
This result shows that the CKM matrix is predominantly a rotation around the third axis $s_3$.
Since $V_{CKM}=V_u^{\dagger}V_d$ the combined effect of the up and down rotation matrices should produce
a rotation mostly around the third axis. In principle, for any choice of $V_d$ there is always
a $V_u$ orthogonal matrix that is consistent with $V_{CKM}$, however we will see that the
observed quark mass hierarchy will reduce substantially these choices.

Our objective is to find the conditions on D-brane inspired mass matrices so that they are
consistent with the experimental data. Thus, given the CKM mixing and the predicted form of
the down (up) quark mass texture, we will use the above analysis to determine the form of the
corresponding up (down) quark matrix.  Starting for example with the down quark mass matrix, and the orthogonal
matrix $V_d$ which diagonalizes the down quarks, then we need in this new formalism a convenient
way to express the diagonalizing  matrix of the  up quarks which is given by  $V_u=V_dV_{CKM}^{\dagger}$.
To express in a simple way the multiplication of two arbitrary orthogonal matrices, we use
the formula (\ref{unit}). Obviously, the resulting matrix is also orthogonal, therefore it
can be expressed in terms of a new unit vector and a new angle.
If we identify $\hat a$ with the unit vector related to the diagonalizing
matrix of the up-quarks and $\hat b$ with the unit vector associated to   the
corresponding one  for the down  quarks, we can apply directly the Cayley-Hamilton
formula  for the multiplication $V_u=V_dV_{CKM}^{\dagger}$,  to obtain
\begin{eqnarray}
\hat a&=& \frac{1}{\sin\frac{\alpha}2}\left\{\sin\frac{\beta}2\cos\frac{\phi_c}2\,\hat b-\cos\frac{\beta}2\,\sin\frac{\phi_c}2\,
\hat n-\sin\frac{\phi_c}2\,
\sin\frac{\beta}2\,\hat b\times\hat n\right\}\label{xxx}
\end{eqnarray}
The angles are also related  by  the additional expression
\ba
\cos\frac{\phi_c}{2}&=&\cos\frac{\alpha}{2}\cos\frac{\beta}{2}+\sin\frac{\alpha}2\,\sin\frac{\beta}2\cos\theta
\ea
where  $\cos\theta=\hat a\cdot\hat b$, thus the CKM angle $\phi_c$ is directly expressed
in terms of the relative declination of the `rotational' axes of the down and up quark
diagonalizing matrices.

Let's see two  limiting cases that will be useful in our subsequent examples:
If we choose $\hat b$ to be aligned with the CKM-axis $\hat b=\hat n$ , then $\hat b\times\hat n=0$ and
$\hat a$ is given as a combination of the two remaining components
on the RHS of (\ref{xxx})
$$\sin\frac{\alpha}2\,\hat a=\left(\sin\frac{\beta}2\cos\frac{\phi_c}2-\cos\frac{\beta}2\,\sin\frac{\phi_c}2\right)\,
\hat n$$
This is satisfied if $\hat a=\pm\hat n$, and
$$\pm\sin\frac{\alpha}2=\sin\frac{\beta-\phi_c}2$$
i.e., if $\beta-\alpha=2k\,\pi+\phi_c$. Choosing $\hat b=-\hat n$, we find $\alpha-\beta=2k\,\pi+\phi_c$.

A second limiting case arises if we choose $\vec a\cdot\vec b=0$. Then $\cos\theta =0$ and the
three angles are related by the simple expression
$$\cos\frac{\phi_c}{2}=\cos\frac{\alpha}2\,\cos\frac{\beta}2.$$

Before closing this section,  let us point out for later convenience that the formula
(\ref{xxx}) can be expressed in a more compact  form.  Indeed, if we redefine
\ba
\hat a&\ra& \vec a=\hat a\sin\frac{\alpha}2\nn\\
\hat b&\ra&\vec b=\hat b\sin\frac{\beta}2\nn
\ea
and
\ba
\hat n&\ra&\vec c=\hat n\sin\frac{\phi_c}2\label{ntoc}
\ea
we rewrite (\ref{xxx}) in the following simplified expression
\begin{eqnarray}
\vec a&=& \cos\frac{\phi_c}2\,\vec b-\cos\frac{\beta}2\,\vec c-\vec b\times\vec c\label{xyz}
\end{eqnarray}

\section{Analytic expressions }

In our previous examples of mass matrices emerging in effective models
 from consistent D-brane configurations, we have seen that novel
 fermion mass  textures appear where their entries do not exhibit
 the expected `hierarchical' pattern.

In fact, for the cases presented previously  the mass entries of a certain line of
the mass matrix emerge from a certain source, thus, each line is characterized by a certain
 mass scale which in general differs from the scale of another line. Clearly,
this scale should be correlated to some  appropriate (eventually invariant under some specific operation)
 quantity and {\it not} to a single coupling which might be accidentally zero  due to some particular
 choice of the basis.   In this case in our subsequent analysis it is adequate to deal mainly with the magnitudes
of the vectors $\vec \xi_j^{u,d}$, $j=1,2,3$, or equivalently the magnitudes of the lines of the
corresponding triangular matrix (\ref{Chol}).

We can determine all possible acceptable mass textures if  we  express the elements of
the corresponding triangular mass matrices only as functions of the directional
cosines and the mass eigenvalues, i.e., the masses of the quarks.

To proceed further we recall from the analysis of the previous sections that for a given
(in general) non-symmetric matrix $m$, it holds $mm^T=M_CM_C^T$, where $M_C$ is the Cholesky from.
In the present section we give analytic formulae for the triangular form $M_C$
with its entries  expressed only in terms of the eigenmasses of $mm^T$  and the
elements of its diagonalizing orthogonal matrix. Let us start with the evaluation of the entries
of the orthogonal transformation.  Assuming  the orthogonal  transformation $U(\alpha)$  given
by the Cayley-Hamilton formula while redefining
\begin{equation}
\widehat{a}\rightarrow \sin \frac{\alpha}{2}\,\vec{a}~\cdot
\end{equation}%
 we get the following simplified expression for the orthogonal matrix
\ba
U(\alpha)=1+2\cos \frac{\alpha}{2}\overrightarrow{a}\cdot\overrightarrow{s}+2\left(
\overrightarrow{a}\cdot\overrightarrow{s}\right) ^{2}\label{UniV}
\ea
where now the vector $\vec a$ is no logger a unit vector,
\ba
\overrightarrow{a}^{2}&=&a_{1}^{2}+a_{2}^{2}+a_{3}^{2}=\sin ^{2}\frac{\alpha}{2}\label{avec}
\ea
It is useful to write the orthogonal matrix in  expanded form. We get
\begin{equation}
U({\alpha})=%
\begin{bmatrix}
\cos \alpha+2a_{1}^{2} & 2\left( a_{3}\cos \frac{\alpha}{2}-a_{1}a_{2}\right)  &
2\left( a_{2}\cos \frac{\alpha}{2}+a_{1}a_{3}\right)  \\
-2\left( a_{3}\cos \frac{\alpha}{2}+a_{1}a_{2}\right)  & \cos \alpha+2a_{2}^{2} &
2\left( a_{1}\cos \frac{\alpha}{2}-a_{2}a_{3}\right)  \\
-2\left( a_{2}\cos \frac{\alpha}{2}-a_{1}a_{3}\right)  & -2\left( a_{1}\cos \frac{%
\alpha}{2}+a_{2}a_{3}\right)  & \cos \alpha+2a_{3}^{2}%
\end{bmatrix}\label{abc}
\end{equation}%
where $a_{1,2,3}$ are not all independent since they satisfy (\ref{avec}).  Therefore,
$U(\alpha)$ is expressed  only in terms of three independent parameters as expected.
We can now express analytically the elements of the triangular matrix as functions
of the orthogonal matrix entries $u_{ij}$ and the mass eigenstates as follows
\ba
\overrightarrow{\xi}_{1}\cdot \overrightarrow{e}_{1}&=&\sqrt{%
u_{11}^{2}m_{1}^{2}+u_{12}^{2}m_{2}^{2}+u_{13}^{2}m_{3}^{2}}
\nn\\
\overrightarrow{\xi}_{2}\cdot \overrightarrow{e}_{2}&=&\sqrt{\frac{%
u_{33}^{2}m_{1}^{2}m_{2}^{2}+u_{31}^{2}m_{2}^{2}m_{3}^{2}+u_{32}^{2}m_{1}^{2}m_{3}^{2}%
}{u_{11}^{2}m_{1}^{2}+u_{12}^{2}m_{2}^{2}+u_{13}^{2}m_{3}^{2}}}
\nn\\
\overrightarrow{\xi}_{3}\cdot \overrightarrow{e}_{3}&=&\frac{m_{1}m_{2}m_{3}}{%
\sqrt{%
u_{33}^{2}m_{1}^{2}m_{2}^{2}+u_{31}^{2}m_{2}^{2}m_{3}^{2}+u_{32}^{2}m_{1}^{2}m_{3}^{2}%
}}
\nn\\
\overrightarrow{\xi}_{2}\cdot \overrightarrow{e}_{1}&=&\frac{%
u_{11}u_{21}m_{1}^{2}+u_{22}u_{12}m_{2}^{2}+u_{13}u_{23}m_{3}^{2}}{\sqrt{%
u_{11}^{2}m_{1}^{2}+u_{12}^{2}m_{2}^{2}+u_{13}^{2}m_{3}^{2}}}\label{Cip}
\\
\overrightarrow{\xi}_{3}\cdot \overrightarrow{e}_{1}&=&\frac{%
u_{11}u_{31}m_{1}^{2}+u_{32}u_{12}m_{2}^{2}+u_{33}u_{13}m_{3}^{2}}{\sqrt{%
u_{11}^{2}m_{1}^{2}+u_{12}^{2}m_{2}^{2}+u_{13}^{2}m_{3}^{2}}}
\nn\\
\overrightarrow{\xi}_{3}\cdot \overrightarrow{e}_{2}&=&-\frac{%
u_{23}u_{33}m_{1}^{2}m_{2}^{2}+u_{21}u_{31}m_{2}^{2}m_{3}^{2}+u_{22}u_{32}m_{1}^{2}m_{3}^{2}%
}{\sqrt{u_{11}^{2}m_{1}^{2}+u_{12}^{2}m_{2}^{2}+u_{13}^{2}m_{3}^{2}}\sqrt{%
u_{33}^{2}m_{1}^{2}m_{2}^{2}+u_{31}^{2}m_{2}^{2}m_{3}^{2}+u_{32}^{2}m_{1}^{2}m_{3}^{2}%
}}~.\nn
\ea
Thus, the triangular matrix elements are simple functions of the eigenmasses and the
orthogonal transformation entries. This analytic result simplifies remarkably the analysis
of classifying experimentally admissible mass matrices, while we note that several simple
textures can  be found even  by  simple inspection of the above analytic structure.

 We can use an alternative parametrization of the triangular matrix reliant on the relations
 (\ref{Cip}) as follows. We define the diagonal matrix of the squared  mass eigenvalues
\ba
{\cal M}&=&\left(\begin{array}{ccc}
m_1^2&0&0\\
0&m_2^2&0\\
0&0&m_3^2\end{array}\right)
\ea
and the vectors
\ba
\vec v_i&=&(u_{i1},u_{i2},u_{i3})\label{vecs}
\ea
where $u_{ij}$ are the elements of the diagonalizing matrix, thus $\vec v_i\cdot \vec v_j=\delta_{ij}$.
In this notation, the nominator of the $\{21\}$ entry of the triangular matrix is written
\ba
\vec\xi_2\cdot\hat e_1&\propto&\vec v_2{\cal M}\vec v_1\;\equiv\;\vec v_1{\cal M}\vec v_2=\sum_{i=1}^3u_{2i}m_i^2u_{1i}
\ea
Similarly, we find also also that the  $\{31\}$ entry is proportional to
\ba
\vec\xi_3\cdot\hat e_1&\propto&\vec v_3{\cal M}\vec v_1\;\equiv\;\vec v_1{\cal M}\vec v_3=\sum_{i=1}^3u_{3i}m_i^2u_{1i}
\ea
and the  $\{32\}$
\ba
\vec\xi_3\cdot\hat e_2&\propto&\vec v_3{\cal M}^{-1}\vec v_2\;\equiv\;\vec v_2{\cal M}\vec v_3=\sum_{i=1}^3u_{3i}m_i^{-2}u_{2i}
\ea
In this notation, all entries are expressed in terms of inner products $\vec v_j{\cal M}^n\vec v_i$, $n={\pm 1}$,
and the triangular mass matrix takes the following elegant form
\ba
M_C&=&\left(\begin{array}{ccc}
 \sqrt{\vec v_1{\cal M}\vec v_1}&0&0\\
   \frac{\vec v_2{\cal M}\vec v_1}{\sqrt{\vec v_1{\cal M}\vec v_1}}&\sqrt{\frac{\vec v_3{\cal M}^{-1}\vec v_3}{\vec v_1{\cal M}\vec v_1}}\,m_1m_2m_3&0\\
    \frac{\vec v_3{\cal M}\vec v_1}{\sqrt{\vec v_1{\cal M}\vec v_1}}&
     - \frac{\vec v_3{\cal M}^{-1}\vec v_1\,m_1m_2m_3}{\sqrt{(\vec v_1{\cal M}\vec v_1)(\vec v_3{\cal M}^{-1}\vec v_3)}}&\frac{1}{\sqrt{\vec v_3{\cal M}^{-1}\vec v_3}}\\
\end{array}
\right)\label{vecM}
\ea
The form (\ref{vecM}) of the triangular matrix will prove particularly
 useful for the classification of the textures with zeroes discussed in the subsequent sections.

\subsection{Mass textures with zeroes}

In phenomenological investigations, a usual practice to minimize the number of arbitrary
mass parameters to those which suffice to determine  the quark mass eigenstates, is to seek viable
texture-zero mass matrices. In this section we are going to explore in detail this issue
motivated also by the fact that in several cases  of String and D-brane models, negligible or even zero
 entries in Yukawa textures do persist even after the inclusion of  non-renormalizable
 or other contributions, because of  remnant discrete or other symmetries left over  from the higher theory.
 This is also the case in F-theory constructions when some matter fields are localized
on different curves~\cite{Beasley:2008kw}.

The analytic result obtained above for the Cholesky form of a matrix allows the classification of texture-zero
mass matrices  in a simple and elegant way. We first note that the Cholesky form of the matrix
is already a non-symmetric texture-zeroes Yukawa matrix  itself. Using suitable values for the angles $(\alpha,\beta,\gamma)$
of the orthogonal transformation (\ref{Ortho})  we can obtain more  texture-zero forms of the mass matrix
$m_D$.  Up to possible signs, while without assuming  any  further relation of the diagonalizing matrix
and the mass matrix entries,  we find that in addition to (\ref{Chol})  there are only four more non-symmetric
texture zero forms, namely
\ba
\left(
\begin{array}{ccc}
 0 & 0 & a_1 \\
 0 & -b_2 & b_1 \\
 c_3 & -c_2 & c_1
\end{array}
\right)&,&\left(
\begin{array}{ccc}
 0 & 0 & a_1 \\
 -b_2 & 0 & b_1 \\
 -c_2 & -c_3 & c_1
\end{array}
\right)\nn\\
\left(
\begin{array}{ccc}
 0 & a_1 & 0 \\
 0 & b_1 & b_2 \\
 c_3 & c_1 & c_2
\end{array}
\right)&,&\left(
\begin{array}{ccc}
 0 & a_1& 0 \\
 -b_2 & b_1 & 0 \\
 -c_2 & c_1 & c_3
\end{array}
\right)\nn
\ea
These essentially correspond to simple rearrangements of the zeroes of the
initial matrix under trivial transformations. 

 A less trivial and more appealing task
is of course to  minimize further the arbitrary parameters of (\ref{Chol}) by
setting additional entries equal to zero yet reconciling the experimental
data.  To this end, we proceed to a classification of all non-trivial zeroes
of the down-quark mass matrix setting successively in (\ref{Cip}) the off-diagonal
elements equal to zero, i.e. $\vec\xi_j\cdot\hat e_i=0$,  and derive the conditions
implied for the remaining non-zero matrix elements. Having determined the
specific forms of $m_D$, we can use the results of
each individual solution to determine the corresponding up quark texture.
We first start with the observation that none of the three
diagonal elements of the triangular matrix can be set equal to zero since this
would imply that at least one eigenmass is zero which contradicts
the data. Therefore, the only possible zeroes in a Cholesky matrix
can be found in the off-diagonal entries $\{21\},\{31\},\{32\}$.

\subsubsection{The  case of one diagonal quark mass matrix}

We start with the simplest (trivial) possibility where all possible entries
in the down-quark Cholesky mass matrix are zero, i.e. a texture three-zeroes,
$$m_{D_{21}}=m_{D_{31}}=m_{D_{32}}=0$$
This implies that the down quark mass  matrix is in diagonal form, therefore
its diagonalizing matrix is the identity matrix $V_d=I$, whilst $V_u=V_{CKM}^\dagger$.
Therefore the up-quark mass diagonalizing orthogonal matrix has the form (\ref{Chol}),
its elements being $u_{ij}=\left(V_{CKM}^\dagger\right)_{ij}$.
Substituting the appropriate experimental values of masses and mixing, the numerical
form of the matrix is
\begin{eqnarray}
{m_U}&\approx&\left(
\begin{array}{lll}
 0.293 & 0 & 0 \\
 -3.792 & 6.150 & 0 \\
 64.938 & -158.029 & 0.342
\end{array}
\right)
\end{eqnarray}
with all entries expressed in {\rm GeV}. It is easy to see that the square roots of the eigenvalues
of the matrix $m_Um_U^T$ are the up quark masses $m_u= 0.003, m_c=1.21,m_t=171${\rm GeV} as expected.
Thus, in the basis where the down quark mass matrix is diagonal, all entries of the up-quark
mass matrix are non-zero, whilst the line-vectors $\vec\xi_j^u$ (in the notation (\ref{Mvec}) of the
matrix $m_U$) exhibit a hierarchical pattern~\footnote{Because of the property (\ref{magn}) the magnitude of
the `line-vector' $\vec\xi_i$ in (\ref{Mvec}) is equal to that of the corresponding line in
the Cholesky matrix (\ref{Chol}).}
in the sense that $|\vec\xi_1^u|<|\vec\xi_2^u|<|\vec\xi_3^u|$ and this is true for a whole class
of equivalent matrices which are obtained when orthogonal transformations are acting on the
progenitor from the right. Indeed, multiplying by any orthogonal matrix from the RHS, the measures of the
vectors $|\vec\xi_j|$ do not change. For example, acting with an orthogonal transformation the first
vector $\vec\xi_1$ becomes
\begin{eqnarray}
\vec \xi'_1=\{1.52781 \cos (\alpha ) \cos (\gamma ),1.52781 \cos (\gamma ) \sin (\alpha ),1.52781 \sin (\gamma )\}
\end{eqnarray}
while it can be checked that $|\vec\xi_1|=|\vec\xi'_1|$ and in the same manner $|\vec\xi_{2,3}|=|\vec\xi'_{2,3}|$.
Thus, if the down quark mass matrix is cast to diagonal
form, the mass hierarchy and CKM imply definite hierarchical structure
$$|\vec\xi_1|\,:\,|\vec\xi_2|\,:\,|\vec\xi_3|\,\sim \, \rho^4\,:\,\rho^2\,:\,1$$
with $\rho\sim 0.2$. We will see in the next  sections
that  the hierarchy of the vectors $|\vec \xi_i|$ can be reversed  if both matrices are non-diagonal
but in limited regions of the parameter space in order to achieve consistency with the CKM mixing.

\subsubsection{Textures with two-zeros}

We turn now to the non-trivial texture-zero cases with non-diagonal mass matrices. We demand that
the entry $\{21\}$ is zero, thus  we take $\vec\xi_2\cdot \hat e_1=0$ which implies
\ba
u_{11}u_{21}m_1^2+u_{22}u_{12}m_2^2+u_{13}u_{23}m_3^2&=&0
\ea
We first try to satisfy the above condition without assuming a particular relation
between $u_{ij}$ and  the mass eigenvalues $m_i$.  Two possible solutions are
\begin{eqnarray}
i)&&u_{12}=u_{21}=u_{13}=0\;{\rm or}\\
ii)&&u_{12}=u_{21}=u_{23}=0
\end{eqnarray}
Starting with the first case, from $u_{12}=u_{21}=0$ it follows from (\ref{abc})
$$a_3\cos\frac{\alpha}2\pm a_1a_2=0$$
{\it A)}\; A simple (although not the only) way to satisfy the above
 is by choosing $a_2=a_3=0$, so that $a_1=\sin\frac{\alpha}2$ and
 the diagonalizing and Cholesky matrices assume the simplified form
\begin{eqnarray}
U_a=\left(\begin{array}{ccc}
1&0&0\\
0&\cos\,\alpha&\sin\,\alpha\\
0&-\sin\,\alpha&\cos\,\alpha
\end{array}\right)&&M_{C}=\left(\begin{array}{ccc}
x_{11}&0&0\\
0&x_{22}&0\\
0&x_{32}&x_{33}
\end{array}\right)
\end{eqnarray}
with $x_{11}= m_1$, $x_{22}=\pm\sqrt{m_3^2-\delta_{32}^2\cos^2\alpha}, \,x_{33}=m_2m_3/x_{22}$,
and $x_{32}=\delta_{32}^2 \sin(2\alpha)/(2x_{22})$, where $\delta_{ji}^2=(m_j^2-m_i^2)$.

{\it B)}\; If we take the second case, then $a_1=a_3=0$, so that $a_2=\sin\frac{\alpha}2$ and
the matrices are reduced to
\begin{eqnarray}
U_a=\left(\begin{array}{ccc}
\cos\,\alpha&0&\sin\,\alpha\\
0&1&0\\
-\sin\,\alpha&0&\cos\,\alpha
\end{array}\right)&&M_{C}=\left(\begin{array}{ccc}
x_{11}&0&0\\
0&x_{22}&0\\
x_{31}&0&x_{33}
\end{array}\right)
\end{eqnarray}
with $x_{11}=\sqrt{m_3^2-\delta_{31}^2\cos^2\alpha},x_{22}=m_2,x_{33}=m_1m_3/x_{11}$, and
$x_{31}=\delta_{31}^2\sin(2\alpha)/(2x_{11})$.

{\it C)}\; Next we  assume that $\vec\xi_3\cdot \hat e_1=0$. This implies one new case,
$a_1=a_2=0$ and the following structure
\begin{eqnarray}
U_a=\left(\begin{array}{ccc}
\cos\,\alpha&\sin\,\alpha&0\\
-\sin\,\alpha&\cos\,\alpha&0\\
0&0&1
\end{array}\right)&&M_{C}=\left(\begin{array}{ccc}
x_{11}&0&0\\
x_{21}&x_{22}&0\\
0&0&x_{33}
\end{array}\right)
\end{eqnarray}
with $x_{11}=m_2^2-\delta_{21}^2\cos\alpha$, $x_{22}=\pm m_1m_2/x_{11}, x_{33}=m_3$ and $x_{21}=\delta_{21}^2\sin(2\alpha)/(2x_{11})$.

A thorough and  rigorous analysis of the textures with two-zeroes is presented in detail
in the appendix. It is shown that all other possible solutions can be reduced to the
 above three cases by trivial transformations.
We observe that  when no specific relation between the eigenvalues  $m_i$ and
the mixing entries $u_{ij}$ is assumed,
the triangular matrix cannot have a single zero off-diagonal element. There are
always two zeroes in the triangular matrix which are always
correlated to one of the three  angles in the  diagonalizing matrix (\ref{Ortho}).

For any of the above cases we can now compute the diagonalizing up-quark orthogonal
matrix and use this result to calculate the entries of the corresponding  up quark
mass matrix. All elements of the two triangular matrices can be expressed in terms
 of  one free parameter, namely the angle $\alpha$.   In deploying our procedure
 in the next subsection we are going to use this freedom to pin down quark mass patterns,
in particular those which can be in accordance with specific classes
from D-brane configurations.

\subsubsection{Textures with one zero}

 The case of triangular matrices with only one zero element is more involved. We have seen
in the previous subsection that whenever we demand one of the entries to be zero, there is always
a second zero element in the matrix unless a non-trivial relation between $u_{ij}$ and
the mass eigenstates $m_i$ is imposed.  In the following, we will derive and discuss in detail
these conditions for the case that the only zero is  $\vec\xi_2\cdot\hat e_1=0$. The analysis can be easily
extended  to the other two non-diagonal elements of the triangular matrix and for
completeness is presented in the appendix.  To proceed, we
use the vector-like formalism of the triangular matrix elements introduced in (\ref{vecM}).
Then, the $\{21\}$ entry of the triangular matrix is proportional to
\ba
\vec\xi_2\cdot\hat e_1&\propto&\vec v_2{\cal M}\vec v_1\;\equiv\;\vec v_1{\cal M}\vec v_2=\sum_{i=1}^3u_{2i}m_i^2u_{1i}
\ea
The requirement that the  $\{21\}$ element in the mass matrix is equal to zero,  is now equivalent to the orthogonality condition  $$\vec v_2{\cal M}\vec v_1=\vec v_1{\cal M}\vec v_2=0$$
The condition $\vec v_2{\cal M}\vec v_1=0$  implies that the vector ${\cal M}\vec v_1$ is orthogonal to $\vec v_2$ and therefore can be expressed
as a  linear combination of $\vec v_1,\vec v_3$. Similarly $\vec v_1{\cal M}\vec v_2=0$ implies that
${\cal M}\vec v_2$ can be expressed in terms of $\vec v_2,\vec v_3$. We find
\ba
{\cal M}\vec v_1&=& \left(\vec v_1{\cal M}\vec v_1\right)\,\vec v_1+\left(\vec v_3{\cal M}\vec v_1\right)\,\vec v_3
\nn\\
{\cal M}\vec v_2&=& \left(\vec v_2{\cal M}\vec v_2\right)\,\vec v_2+\left(\vec v_3{\cal M}\vec v_2\right)\,\vec v_3
\nn
\ea
Both of the above can be solved for $\vec v_3$, giving
\ba
\vec v_3&=&\frac{1}{\left(\vec v_3{\cal M}\vec v_k\right)}\left({\cal M}- \vec v_k{\cal M}\vec v_k\right)\,\vec v_k,\;k=1{\rm or}\; 2
\ea
or, in component form ($A_{ji}=\vec v_j{\cal M}\vec v_i=\sum_{i=1}^3u_{jl}m_l^2u_{il}$)
\ba
u_{3j}&=&\frac{m_j^2-A_{kk}}{A_{3k}}
\,u_{kj},\;k=1,2,\;j=1,2,3\label{zero1tex}
\ea
Thus, the relations (\ref{zero1tex}) are sufficient to ensure that at least the $\{21\}$ element of the triangular mass matrix is zero, while similar relations hold for the vanishing of the other off-diagonal elements. By simple inspection of the above
formula we find that  if some  $u_{kj}$  on the right-hand side of (\ref{zero1tex}) is set equal to zero, then
 the  two-zeroes textures   discussed previously are recovered. Indeed, to make clear the argument,
let us explore a particular case setting $u_{13}=0$. Putting $k=1$ in the above, we find that this implies $u_{33}=0$, while due to $u_{13}^2+u_{23}^2+u_{33}^2=1$ we get $u_{23}=1$. Putting $k=2$, we get
\ba
(m_3^2-A_{22})\,u_{23}&=&0
\nn
\ea
Since  $u_{23}$ is non-zero, ($u_{23}=1), $ we get
\ba
m_3^2&=&A_{22}=u_{21}^2m_1^2+u_{22}^2m_2^2+u_{23}^2m_3^2
\ea
which imposes the additional condition
\ba
u_{21}^2m_1^2+u_{22}^2m_2^2&=&0\nn
\ea
i.e., $u_{21}=u_{22}=0$. We further find $\vec\xi_3\cdot\hat e_2=0$, thus this case is reduced to textures with two-zeroes.

In the same way, we can prove that if any of the elements $u_{ki}$, $k=1,2, i=1,2,3$ is set equal to zero,
we arrive at one of the textures with two-zeroes discussed previously.

 Therefore,  distinct, texture zero-one cases are possible only when all $u_{kj}\ne 0$,
 with $k$ taking the values 1 or 2 as above. This of course does not exclude that
 some of the remaining entries $u_{3j}$ could not be  zero.  In the next section we
 will present one such simple example of one zero  texture mass matrix which is also
 compatible with D-brane patterns discussed in section 3.

\subsection{On the relation with the symmetric texture-zero matrices}

It is worth exploring the connection of the above triangular texture-zero analysis with
the symmetric texture-zeroes already discussed in the literature sometime ago~\cite{Ramond:1993kv}.
Using the mathematical analysis presented in previous section, it is straightforward to
 bring the latter into their corresponding triangular form. This calculation shows that
 from the set of the five texture zero symmetric mass matrices only one pair can be identified
 with a two-zeroes texture and one more with a the one-zero texture  of our analysis.
  In particular, we find the following  texture-zero triangular form
\begin{equation}
m_U=%
\begin{bmatrix}
\varepsilon ^{6} & 0 & 0 \\
0 & \varepsilon ^{2}\sqrt{1+\varepsilon ^{8}} & 0 \\
\varepsilon ^{2} & \frac{1}{\sqrt{1+\varepsilon ^{8}}} & \frac{\varepsilon
^{4}}{\sqrt{1+\varepsilon ^{8}}}%
\end{bmatrix}%
,m_D=%
\begin{bmatrix}
2\varepsilon ^{4} & 0 & 0 \\
2\varepsilon ^{3} & 2\varepsilon ^{3}\sqrt{1+\varepsilon ^{2}} & 0 \\
2\varepsilon ^{3} & \frac{1}{\sqrt{1+\varepsilon ^{2}}} & \frac{\varepsilon
}{\sqrt{1+\varepsilon ^{2}}}%
\end{bmatrix}%
\end{equation}
which can be easily converted to one of the symmetric textures using the analysis of section 2.

 The remaining three textures of ~\cite{Ramond:1993kv} -from the point of view of their progenitors-
 correspond to the general form with non-zero off-diagonal entries and the zeroes appear only in a certain
 point of the parameter space. To make this point clear, let us define the following triangular
mass matrix for the  down quarks
\ba
m_D&=&\left(
\begin{array}{lll}
 2 \epsilon ^4 & 0 & 0 \\
 2 \epsilon ^3 & 2 \epsilon ^3 \sqrt{\epsilon ^2+4} & 0 \\
 4 \epsilon ^3 & \frac{2}{\sqrt{\epsilon ^2+4}} & \frac{\epsilon }{\sqrt{\epsilon ^2+4}}
\end{array}
\right)
\ea
Diagonalizing   the symmetric $m_Dm_D^T$, while choosing $\epsilon\sim .23$,
we find   the mass eigenvalues in good agreement
with the values $m_d,m_s,m_b$.
According to our discussion, there is an equivalent class of mass matrices obtained by
the following $U$-action on $m_D$, $m_D\,U$, where $U$ is any orthogonal matrix. If we restrict to the orthogonal
matrices, $U$ this is given by (\ref{Ortho}), thus the parameter space is defined by
the angles $\alpha,\beta,\gamma$. Making the particular choice
\ba
\alpha_0=\frac{\pi}2,\beta_0=\cos ^{-1}
\left(-\frac{\epsilon }{\sqrt{\epsilon ^2+4}}\right),\gamma_0=0\label{choice0}
\ea
we construct the equivalent symmetric texture-zero form
\ba
M_D=m_D\,{U}(\alpha_0,\beta_0,\gamma_0)=\left(
\begin{array}{ccc}
 0 & 2 \epsilon ^4 & 0 \\
 2 \epsilon ^4 & 2 \epsilon ^3 & 4 \epsilon ^3 \\
 0 & 4 \epsilon ^3 & 1
\end{array}
\right)\label{aczero}
\ea
which is one of the texture-zeroes down quark mass matrices proposed in \cite{Ramond:1993kv}.
Therefore, the zeroes in (\ref{aczero}) are completely accidental and arise due to
the particular choice (\ref{choice0}).  An infinite number of equivalent mass matrices implies the
same `physical' quantities, namely the mass eigenstates and the diagonalizing matrix.
 For completeness, we give in the appendix the triangular form of the remaining four
 texture-zero symmetric quark mass matrices.

\section{Examples of Admissible D-brane textures}

Our present analysis has been motivated by the peculiar patterns of mass matrices
which have appeared in certain D-brane configurations accommodating the Standard Model
gauge symmetry. In this section, we will give simple examples where some of these
D-brane inspired textures can appear, at least in some regions of the parameter space.
Of course, the possibilities of finding consistent textures do increase if we assume the most
general triangular mass matrices without imposing the rather restrictive conditions for zeroes.
However, a complete analysis is beyond the scope of this paper. Instead, our aim is to
illustrate how the new formalism applies in representative examples. We will concentrate
in the case of texture-zero triangular forms and try to find some of the D-brane inspired
textures that reconcile the experimental data.

\subsection{The two-zeroes case}
Once we have determined the texture with two-zeroes  and the diagonalizing matrix of the
down quarks, we can use the relations given in the previous section to construct
the corresponding up quark mass texture. As an example of the method, we
start with the down quark matrix such that the rotation is around the first axis, so the
vector is $\vec b=(b_1,0,0)$, with $b_1=\sin\frac{\beta}2$. The diagonalizing matrix is
\ba
\left(
\begin{array}{ccc}
 1 & 0 & 0 \\
 0 & \cos \beta & \sin \beta \\
 0 & -\sin \beta & \cos \beta
\end{array}
\right)
\ea
and the down quark matrix is found to be
\ba
m_D&=&\left(
\begin{array}{ccc}
 m_d & 0 & 0 \\
 0 & \sqrt{m_s^2 \cos ^2\beta+m_b^2 \sin ^2\beta}& 0 \\
 0 & \frac{(m_b^2-m_s^2) \sin (2\beta)}{2\sqrt{m_s^2 \cos ^2\beta+m_b^2 \sin ^2\beta}} & \frac{m_b
  m_s}{\sqrt{m_s^2 \cos ^2\beta+m_b^2 \sin ^2\beta}}
\end{array}
\right)\nn
\ea
i.e., a two-zeroes triangular texture as expected. Using (\ref{xyz}) we can determine the vector
components of the diagonalizing matrix $V_u$
\ba
a_1&=&-\cos\frac{\beta}2\,c_1+\cos\frac{\phi_c}2\,\sin\frac{\beta}2\label{upa1}\\
a_2&=&-\cos\frac{\beta}2\,c_2+\sin\frac{\beta}2\,c_3\label{upa2}\\
a_3&=&-\cos\frac{\beta}2\,c_3-\sin\frac{\beta}2\,c_2\label{upa3}
\ea
where the numerical values of the components $c_i,i=1,2,3$ are calculated using (\ref{ntoc})
and (\ref{nCKM}).
First, we consider the particular case where the up-quark mass
matrix is rotated on the orthogonal direction, i.e., we demand  the vector $\vec a$ to take
the form $\vec a=(0,a_2,a_3)$. Imposing $a_1=0$ while using (\ref{xyz}) we find
\ba
\tan\frac{\beta}2&=&+\frac{c_1}{\cos\frac{\phi_c}2}\equiv n_1\,\tan\frac{\phi_c}2\label{ab01}\\
a_2&=&-\frac{\cos\frac{\beta}2}{\cos\frac{\phi_c}2}\,\left\{c_2\cos\frac{\phi_c}2-c_1c_3\right\}\label{ab02}\\
a_3&=&-\frac{\cos\frac{\beta}2}{\cos\frac{\phi_c}2}\,\left\{c_3\cos\frac{\phi_c}2+c_1c_2\right\}\label{ab03}
\ea
The angle entering the up-quark diagonalizing matrix is also fixed in this case and given by
\ba
\sin\frac{\alpha}2&=&\sqrt{\sin^2\frac{\phi_c}2-c_1^2}\equiv \sin\frac{\phi_c}2\sqrt{1-n_1^2}
\label{ab0}
\ea
Equations (\ref{ab02}-\ref{ab0}) determine completely all the entries of the up-quark mass matrix.
Therefore, in the case of  diagonalizing matrices which fulfil the `orthogonality condition'
$\vec a\cdot\vec b= 0$, all entries of the  up and down quark triangular mass matrices are
completely determined. From (\ref{ab0})
we deduce that the angle $\alpha$ is of the order of the Cabbibo angle and therefore the
elements of the up-quark matrix exhibit also a hierarchical structure in the sense described above.

To examine the more general case  we relax the condition $a_1=0$, therefore
(\ref{ab01}-\ref{ab02}) are no longer  valid in this case. The appropriate formulae
for the elements of the vector $\vec a$ are given by (\ref{upa1}-\ref{upa3}). The two angles
$\alpha, \beta$ of the up and down diagonalizing matrices are connected through the relation
\ba
a_1^2+a_2^2+a_3^2&=&\sin^2\frac{\alpha}2
\ea
with $a_i$ given by (\ref{upa1}-\ref{upa3}) thus, the only free parameter is the down quark
angle $\beta$. It can be checked that for arbitrary values of the free parameter $\beta$
all entries of the triangular up-quark mass matrix are non-zero, whenever the corresponding
down quark texture has two zeroes as in the present case. Texture zeros for the up-quarks
in this case are obtained only for specific values of the angle $\beta$.

\begin{figure}[h]
\centering
\includegraphics[scale=.8]{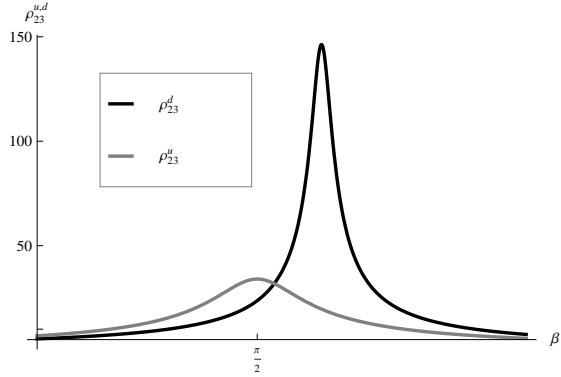}
\caption{The ratios $\rho_{23}^{u,d}$ of the two magnitudes  $|\vec\xi_2^{u,d}|=\sqrt{m_{2j}^2}$ and $|\vec\xi_3^{u,d}|=\sqrt{m_{3j}^2}$
for the up (gray curve) and down quark (black curve) triangular matrices as a function
of the angle $\beta$.  }
 \label{r32}
\end{figure}

Next,  we define the ratios $\rho_{23}=\frac{|\vec\xi_2|}{|\vec\xi_3|}$ for the up
and down  triangular quark mass matrices which, in terms of the triangular mass
matrix entries are given by
\ba
\rho_{23}^{u,d}&=&\sqrt{\frac{{(m_{21}^{u,d})}^2+{(m_{22}^{u,d})}^2}{{(m_{31}^{u,d}})^2+{(m_{32}^{u,d})}^2
+{(m_{33}^{u,d})}^2}}\nn
\ea
These are plotted in figure \ref{r32} as a function of the only free parameter, namely the angle $\beta$.
For  small angle regions,  $\beta\le \frac{\pi}4$ both textures exhibit a hierarchy
$|\vec\xi_2|\le |\vec\xi_3|$ which is reversed for large values of $\beta$.
Furthermore, we observe that textures with two-zeroes have the tendency to be aligned, so up and down
quark mass matrices show the same hierarchy in their `vector like' pattern.  For  small ranges
around $\beta\sim\frac{\pi}4$ the two vectors have comparable magnitudes. This latter case
could fit the first set of  textures obtained in our D-brane scenario, if for example we arrange
the Higgs vevs so that $\kappa_d\gg 1$ and $\kappa_u\ll 1$. We also observe that using the
free parameter $\beta$ we can  obtain zero textures for the up-quark matrix by demanding that
some of the off-diagonal $m_U$-entries are zero.

\begin{figure}[h]
\centering
\includegraphics[scale=.6]{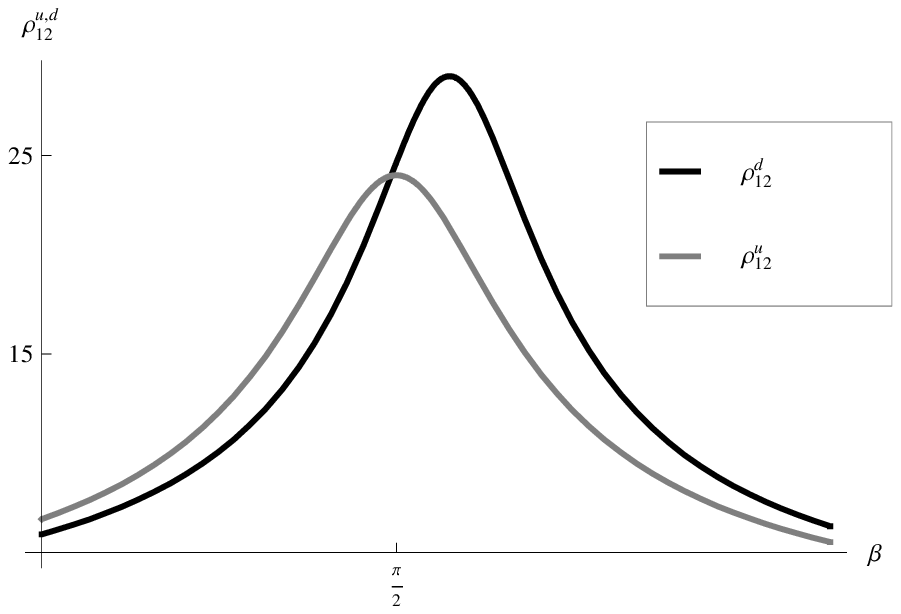}
\includegraphics[scale=.6]{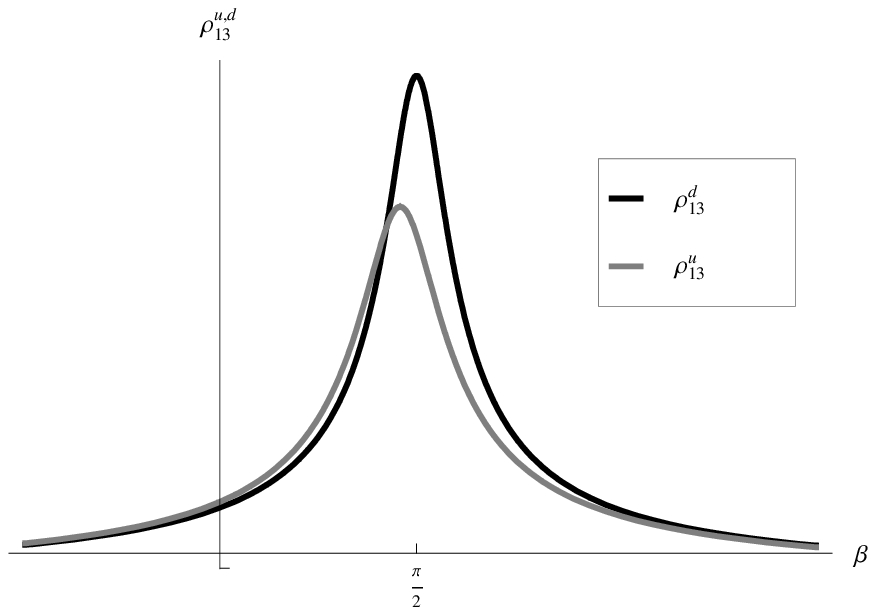}
\caption{The ratios $\rho_{12}^{u,d}$  and $\rho_{13}^{u,d}$  as in figure \ref{r32}.}
 \label{r1213}
\end{figure}

We may elaborate the remaining two cases of the  down quark textures with  two-zeroes and
obtain structures similar to the other two mass matrices obtained in the D-brane
construction of section 3. In Table~\ref{2ztABC} examples of texture-zeroes are presented
for all three cases and in figure \ref{r1213} the ratios similar to those of fig\ref{r32} are
depicted.
\begin{table}[!t]
\centering
\renewcommand{\arraystretch}{1.2}
\begin{tabular}{lccc}
Case&  $\beta$&$m_U$&$m_D$\\
  \hline
$A$
&$\frac{\pi}{1.952}$ &$\left(
\begin{array}{ccc}
 0.293 & 0 & 0 \\
 65.036 & 158.149 & 0 \\
 1.264 & 0 & 0.0134
\end{array}
\right)$&$\left(
\begin{array}{ccc}
 0.005 & 0 & 0 \\
 0 & 4.247 & 0 \\
 0 & -0.165 & 0.125
\end{array}
\right)$\\
&$\frac{\pi}{1.928}$ &$\left(
\begin{array}{ccc}
 0.293& 0 & 0 \\
 65.048 & 158.119 & 0 \\
 0 & -3.073 & 0.0133
\end{array}
\right)$&$\left(
\begin{array}{ccc}
 0.005 & 0 & 0 \\
 0 & 4.243 & 0 \\
 0 & -0.248 & 0.125
\end{array}
\right)$\\&  $\frac{\pi}{53.86}$&$\left(
\begin{array}{ccc}
 0.293 & 0 & 0 \\
 0 & 3.073 & 0 \\
 65.048 & 158.118 & 0.684
\end{array}
\right)$&$\left(
\begin{array}{ccc}
 0.005 & 0 & 0 \\
 0 & 0.277 & 0 \\
 0 & 3.786 & 1.915
\end{array}
\right)$\\
\hline
$B$&$\frac{\pi}{1.973}$&$\left(
\begin{array}{lll}
 7.226 & 0 & 0 \\
 0 & 0.249& 0 \\
 -168.576 & 27.791 & 0.342
\end{array}
\right)$&
$\left(
\begin{array}{lll}
 0.120 & 0 & 0 \\
 -0.003 & 0.005 & 0 \\
 0 & 0 & 4.25
\end{array}
\right)$\\
&$\frac{\pi}{200}$&$\left(
\begin{array}{lll}
 0.253 & 0 & 0 \\
 -1.174 & 7.130 & 0 \\
 0 & -170.851 & 0.342
\end{array}
\right)$&$\left(
\begin{array}{lll}
 0.005 & 0 & 0 \\
 0.0420 & 0.112 & 0 \\
 0 & 0 & 4.25
\end{array}
\right)$\\
\hline
$C$&$\frac{\pi}2$&$\left(
\begin{array}{lll}
 170.851 & 0 & 0 \\
 -7.129 & 1.170 & 0 \\
 0 & 0.271 & 0.003
\end{array}
\right)$&$\left(
\begin{array}{lll}
 4.25 & 0 & 0 \\
 0 & 0.12 & 0 \\
 0.003 & 0 & 0.005
\end{array}
\right)$\\
\end{tabular}
\caption{Examples of the three cases $A,B,C$
of the  two-zeroes  down quark (Cholesky) mass textures.  The
specific values of the free parameter $\beta$ result also to up quark textures
with one zero.  $\beta$-angle values around $\frac{\pi}2$ inverse the hierarchy. In particular,
   $|\xi_2^u|>|\xi_3^u|>|\xi_1^u|$ in $A$,
$|\xi_3^u|>|\xi_1^u|>|\xi_2^u|$ in $B$ and $|\xi_1^u|>|\xi_2^u|>|\xi_3^u|$ in $C$.
}
\label{2ztABC}
\end{table}

\subsection{ One-zero Textures}
\begin{figure}[h]
\centering
\includegraphics[scale=.6]{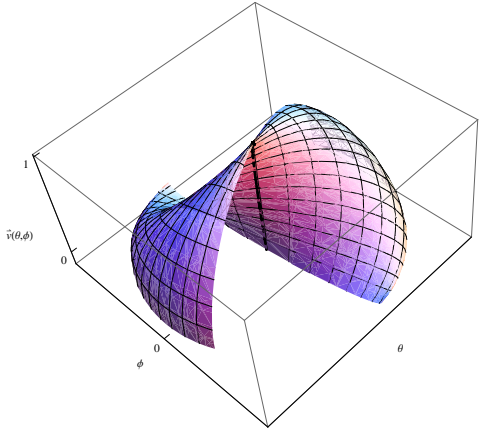}
\caption{Part of the two-dimensional surface spanned by the vector $\vec v_3(\theta,\phi)$. Down quark
masses fix one parameter $\phi=\phi_0$ and the two dimensional surface reduces to the thick curve.}
 \label{rot}
\end{figure}

\begin{figure}[h]
\centering
\includegraphics[scale=.6]{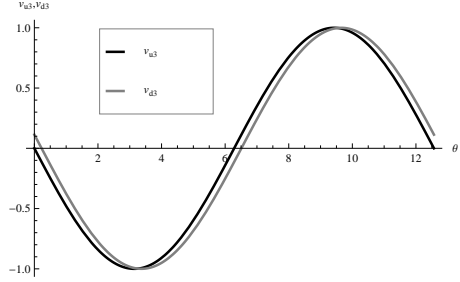}
\caption{ The third components of the vectors of the up and down diagonalizing matrices
 in the representation (\ref{UniV}) as a function of the free parameter in the one-zero texture
 example of the text.}
 \label{vud3}
\end{figure}

\begin{figure}[h]
\centering
\includegraphics[scale=.7]{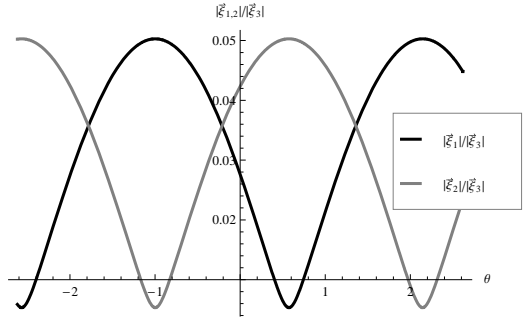}
\caption{The ratios $\frac{|\vec\xi_1|}{|\vec\xi_3|}$ and $\frac{|\vec\xi_2|}{|\vec\xi_3|}$  of the
up-quark ``line-vector'' magnitudes in the one zero texture as function of the free parameter $\theta$.}
 \label{uplot}
\end{figure}

Next, we investigate the consistency conditions  in a simple example with one-zero
texture. We have seen in
the previous section that one-zero triangular mass matrices are possible provided that
certain relations
are imposed between the elements of the diagonalizing and the mass matrices. For example,
imposing (\ref{zero1tex}),
we obtain $\vec\xi_2\cdot\hat e_1=0$ i.e.,   a one-zero texture.  We have further
stressed that to
avoid any other zero off-diagonal entry we must demand $u_{1j}\ne 0$ and $u_{2j}\ne 0 $.
To obtain the simplest
admissible one-zero texture, let us assume however, that  some of the remaining entries of
the orthogonal matrix is zero, i.e., $u_{3j}=0$ for some $j$.
 From (\ref{zero1tex}) we have $m_j^2=A_{kk}$ which implies the relation
 \ba
 u_{kl}^2&=&\frac{m_j^2-m_i^2}{m_l^2-m_j^2}\,u_{ki}^2
 \ea
with $i\ne j\ne l\ne i$ and $k=1,2$. Since $u_{kl}^2, u_{ki}^2$ are always positive, and
the
mass hierarchies are $m_3>m_2>m_1$,  the relation  is valid only for $i=1,j=2,l=3$. Thus
we get
\ba
u_{13}&=&\tan\phi_0\,u_{11}\\
u_{23}&=&\tan\phi_0\,u_{21}
\ea
and
\ba
\tan\phi_0=\sqrt{\frac{m_2^2-m_1^2}{m_3^2-m_2^2}}\label{phi0}
\ea
Working out the details, we find
\ba
V_d&=&\left(
\begin{array}{ccc}
 \cos \theta \cos \phi_0 & -\sin \theta & -\cos \theta \sin \phi_0 \\
 \cos \phi_0 \sin \theta & \cos \theta & -\sin \theta \sin \phi_0 \\
 \sin \phi_0 & 0 & \cos \phi_0
\end{array}
\right)\label{vd1}
\ea
where $\theta$ an arbitrary angle. In the notation of section 5, this can be considered
as a
combined rotation of two orthogonal matrices
\ba
U(\theta)=\exp  [\vec v_{\theta}\cdot\vec s]&,& U(\phi_0)=\exp  [\vec v_{\phi_0}\cdot\vec s]
\ea
with `vectors'
\ba
\vec v_{\theta}=\left(0,-\sin\frac{\theta }{2},0\right)&&
\vec v_{\phi_0}=\left(0,0,-\sin \frac{\phi _0}{2}\right)\nn
\ea
In the same notation, the corresponding `vector' of the combined matrix (\ref{vd1}) is
computed adapting appropriately the formula (\ref{xyz}) for the convolution $U(\theta)U(\phi_0)$
\ba
\vec v_d(\theta,\phi_0)&=&
\left(-\sin \frac{\theta }{2} \sin \frac{\phi _0}{2},-\cos \frac{\theta }{2} \sin
\frac{\phi
   _0}{2},-\sin\frac{\theta }{2}\cos \frac{\phi _0}{2}\right)
\ea
For arbitrary $(\theta,\phi)$ the geometrical locus of the tip of this vector is plotted in figure \ref{rot}.
Given the down quark mass hierarchies and the formula (\ref{phi0}) the angle $\phi$
takes only a specific value $\phi_0$ and the `motion' of the vector tip is constrained
along the curve $\phi=\phi_0$.
We conclude that, unless the free parameter $\theta$ is close to $\theta\sim 2n\pi,
n=0,1,2...$,
the $\vec v_d$ indicates a rotation mainly around the third axis.  The corresponding
`vector' of
the up-quark diagonalizing matrix can be computed using again the very same formula
(\ref{xyz})
where now the combining vectors are the  CKM and $\vec v_d$.
 The third components of both vectors $v_d,v_u$ for $\phi=\phi_0$ are plotted in figure~\ref{vud3}.
The down quark mass matrix is
\ba
m_D&=&\left(
\begin{array}{ccc}
 m_2 & 0 & 0 \\
 0 & m_2 & 0 \\
- \frac{\cos \theta \sqrt{(m_2^2-m_1^2)(m_3^2-m_2^2)} }{m_2} & -\frac{\sin \theta
  \sqrt{(m_2^2-m_1^2)(m_3^2-m_2^2)}}{m_2} & \frac{m_1 m_3}{m_2}
\end{array}
\right)
\ea
For the limiting values $\theta=0,\frac{\pi}2$ of the only free parameter $\theta$
we recover specific patterns of the texture with two zeroes of the previous analysis.
We find the following hierarchy of the vector magnitudes,
$$|\vec\xi_1^d|=|\vec\xi_2^d|=m_2\ll |\vec\xi_3^d|=\sqrt{m_3^2-m_2^2+m_1^2}$$
while as expected they satisfy the relation
$|{\vec\xi_1^d}|^2+|{\vec\xi_2^d}|^2+|{\vec\xi_3^d}|^2=m_1^2+m_2^2+m_3^2$.
 It is straightforward to use the developed formalism and derive the corresponding
analytic expressions for the up-quarks.   It can be checked that consistency with the
experimental data requires also the same hierarchy between the third $|\vec\xi_3^u|$
and the other two $|\vec\xi_{1,2}^u|$ vectors for the up-quark mass matrix, however,
for ranges of $\theta$ the ratio $\frac{|\vec\xi^u_1|}{|\vec\xi^u_2|}$ could be reversed.
To demonstrate this, we plot the ratios $\frac{|\vec\xi^u_1|}{|\vec\xi^u_3|}, \frac{|\vec\xi^u_2|}{|\vec\xi^u_3|}$
in figure \ref{uplot} as a function of the free parameter $\theta$.

In this example, we have dealt with  an one-zero  $m_D$- texture  while we restricted further the investigation
imposing also the condition $u_{32}=0$ on the corresponding diagonalizing matrix.
It is worth seeing also whether  we can minimize the number of parameters in the up-quark mass matrix.
Since we have one free parameter, we might choose an appropriate value to generate a texture zero case for
the up quarks too. Thus, for example, choosing $\theta\sim -\frac{\pi}{3.135}$ we get
\ba
m_U&\approx&
\left(
\begin{array}{lll}
 8.58841 & 0 & 0 \\
 0 & 0.909606 & 0 \\
 -169.892 & 17.458 & 0.0827412
\end{array}
\right)\nn\\
m_D&\approx&\left(
\begin{array}{lll}
 0.12 & 0 & 0 \\
 0 & 0.12 & 0 \\
 -2.30052 & 3.56712 & 0.177083
\end{array}
\right)\nn\\
V_{CKM}&=&\left(
\begin{array}{lll}
 0.974184 & 0.225585 & 0.00877843 \\
 -0.225755 & 0.973331 & 0.0407647 \\
 0.000651602 & -0.0416941 & 0.99913
\end{array}
\right)\nn
\ea
i.e., a one-zero texture,   with zero entry $m_{21}= 0$. The up-quark structure resembles that of case
(\ref{c2}) obtained in the context of D-brane scenarios. The compatibility of the down quark mass matrix
would be possible in the presence of a second Higgs doublet $H_d'$ with appropriate vev so
that  $\kappa_d>1$.

\begin{table}[!t]
\centering
\renewcommand{\arraystretch}{1.2}
\begin{tabular}{ccc}
  $\theta$&$m_U$&$m_D$\\
  \hline
 $-\frac{\pi}{3.135}$ & $ \left(
\begin{array}{lll}
 8.59 & 0 & 0 \\
 0 & 0.91 & 0 \\
 -169.9 & 17.46 & 0.08
\end{array}
\right) $&$\left(
\begin{array}{lll}
 0.12 & 0 & 0 \\
 0 & 0.12 & 0 \\
 -2.30 & 3.57 & 0.177
\end{array}
\right)$\\
&$\left(
\begin{array}{lll}
 1.04 & 0.62 & -8.50 \\
 0.62 & 0.65 & 0.12 \\
 -8.50 & 0.12 & 170.58
\end{array}
\right)$&$\left(
\begin{array}{lll}
 0.09 & 0.05 & -0.07 \\
 0.05 & 0.041 & 0.10 \\
 -0.07 & 0.10 & 4.25
\end{array}
\right)$\\
\end{tabular}
\caption{Case of one-zero up- and down quark (Cholesky) mass textures  $m_{21}^u=m_{21}^d=0$ and
``reversed hierarchy'' for the up-quarks $|\vec\xi_{1,3}^u|>|\vec\xi_2^u|$. Their corresponding
symmetric forms are shown in the second line. }
\label{1zt1}
\end{table}

\begin{table}[!t]
\centering
\renewcommand{\arraystretch}{1.2}
\begin{tabular}{ccc}
  $\theta$&$m_U$&$m_D$\\
  \hline
 $-\frac{\pi}{13.8}$&$\left(
\begin{array}{lll}
 6.20 & 0 & 0 \\
 5.88 & 1.26 & 0 \\
 -170.79 & 0 & 0.08
\end{array}
\right)$&$\left(
\begin{array}{lll}
 0.12 & 0 & 0 \\
 0 & 0.12 & 0 \\
 -4.14 & 0.96 & 0.18
\end{array}
\right)$\\
&$\left(
\begin{array}{lll}
 0.23 & 0.21 & -6.19 \\
 0.21 & 1.461 & -5.83 \\
 -6.19 & -5.83 & 170.58
\end{array}
\right)$&$\left(
\begin{array}{lll}
 0.014 & 0.025 & -0.117 \\
 0.025 & 0.114 & 0.028 \\
 -0.117 & 0.028 & 4.247
\end{array}
\right)$\\
\end{tabular}
\caption{Case of one-zero up- and down quark (Cholesky) mass textures  $m_{32}^u=m_{21}^d=0$ and
`` hierarchy'' for the up-quarks $|\vec\xi_1^u|\sim |\vec\xi_2^u|\ll |\vec\xi_3^u|$. Their corresponding
symmetric forms are shown in the second line.}
\label{1zt2}
\end{table}

\begin{table}[!t]
\centering
\renewcommand{\arraystretch}{1.2}
\begin{tabular}{ccc}
  $\frac{\pi}{5.384}$ & $\left(
\begin{array}{lll}
 0.91 & 0 & 0 \\
 -0.87 & 8.54 & 0 \\
 0 & -170.79& 0.08
\end{array}
\right)$& $\left(
\begin{array}{lll}
 0.12 & 0 & 0 \\
 0 & 0.12 & 0 \\
 -3.54 & -2.34& 0.18
\end{array}
\right)$
\\
  & $\left(
\begin{array}{lll}
 0.66 & -0.63 & -0.03 \\
 -0.63 & 1.02 & -8.50 \\
 -0.03 & -8.50 & 170.58
\end{array}
\right)$&$\left(
\begin{array}{lll}
 0.04 & -0.05& -0.10 \\
 -0.05 & 0.09 & -0.075 \\
 -0.10& -0.07 & 4.25
\end{array}
\right)$ \\
\end{tabular}
\caption{Cases of one-zero mass textures   with $m_{21}^d=0$ and  $m_{31}^u=0$ and their
corresponding symmetric forms.}
\label{1zt3}
\end{table}

All the three possible cases $m_{21}=0,m_{31}=0,m_{32}=0$ can be obtained choosing appropriate
values of the free parameter $\theta$ and are presented respectively in  tables \ref{1zt1},\ref{1zt2} and \ref{1zt3}.
These matrices are ostensibly different, however they encode the same physical properties,
i.e., they result to the same eigenvalues of $m_{U,D}m_{U,D}^T$ and predict the CKM mixing matrix.

\newpage

\section{Conclusions}

In this work we have examined the fermion masses in a wide class of effective low energy models emerging from
intersecting  D-brane configurations where the Yukawa superpotential terms are subject to additional restrictions
from surplus  $U(1)$ symmetries and anomaly cancelation conditions. We have shown that masses for all fermion
 generations are obtained only when additional Higgs doublets, or higher order corrections or substantially
 suppressed non-perturbative effects are taken into account.  We have worked out in detail the spectrum of a representative model with Standard Model gauge symmetry augmented by abelian factors and we have observed
 that the fermion mass matrices exhibit a characteristic pattern which appears in a wide class of models
 obtained in the context of D-brane scenarios.  We investigated specific cases of these novel patterns
 derived for the quark mass  matrices and analyzed in detail the conditions imposed by phenomenological
 constraints on the various mass generating mechanisms  in these constructions. Furthermore, motivated by
 the above considerations, in this work we developed a novel formalism which leads to a unified treatment
 of all viable symmetric and non-symmetric fermion mass textures.  More precisely, we showed that the Cholesky 
 decomposition of the mass matrices captures the features  in a unique way of a whole equivalent class of mass matrices
 which are related to the former by an orthogonal matrix.  The entries of the  corresponding triangular
 (Cholesky) mass  matrix were determined analytically in terms of the eigenmasses and the diagonalizing
 matrix while they were used to explore equivalent classes of symmetric and non-symmetric quark mass matrices
 which reconcile the hierarchical quark mass pattern and the Cabbibo-Kobayashi-Maskawa mixing matrix.
We have further shown that the triangular form of the quark mass matrices admit  texture-zero forms,
minimizing thus the number of arbitrary parameters.  A detailed analysis is presented for the D-brane
derived quark mass patterns and conditions on the various mass generating sources are specified to reconcile
the known mass hierarchies and mixing data.  Finally,  a comparison with the symmetric texture-zero quark
mass matrices existing in the literature is also worked out.

\vfill

 {\it Acknowledgements.}
 This work is partially supported by the European Research and Training Network MRTPN-CT-2006 035863-1
 (UniverseNet).

\newpage

\section{Appendix}

\subsection{The  CKM matrix }

In this appendix we will provide a few more details on the derivation of
 the CKM matrix in the basis $s_1,s_2,s_3$ discussed in section 4.

For the sake of clarity of the our calculations, first  we review in brief the  conventions used
with respect to the diagonalizing matrices of the up and down quarks. We find also
 convenient to adopt here the Wolfenstein  parametrization of the CKM matrix.

Let the weak and mass eigenstates of the up-quarks be related by the orthogonal matrices $V_u^{L,R}$
\ba
u_L^0&=&V_u^L\,u_L\nn\\
u_R^0&=&V_u^R\,u_R\nn
\ea
The relevant Yukawa terms are written
\ba
\bar{u}_L^0m_Uu_R^0&=&\bar{u}_LV_u^{L\dagger}m_UV_u^Ru_R\nn\\
                   &=&\bar{u}_Lm_U^{diag.}u_R
\ea
and similarly for the down quarks. Thus,
the diagonal mass matrices are given in terms of the orthogonal transformations
$V_u^{L,R}$ and $V_d^{L,R}$ respectively as follows
\ba
m_U^{diag.}&=&V_u^{L\dagger}m_UV_u^R\\
m_D^{diag.}&=&V_d^{L\dagger}m_DV_d^R
\ea
Since ${\left(V_u^{L\dagger}m_UV_u^R\right)}^{\dagger}=V_u^{R\dagger}m_U^{{\dagger}}V_u^L$ we get
\ba
\left(m_U^{diag.}\right)^2&=&\left(V_u^{L\dagger}m_UV_u^R\right)\left(V_u^{R\dagger}m_U^{\dagger}{V_u^{L}}\right)
\nonumber\\
&=&V_u^{L\dagger}m_U\,m_U^{\dagger}{V_u^{L}}\ea
and similarly for the down quark matrix$m_D$ . Thus
\ba
m_U\,m_U^{\dagger}&=&{V_u^{L}}\,\left(m_U^{diag.}\right)^2V_u^{L\dagger}\nn\\
m_D\,m_D^{\dagger}&=&{V_d^{L}}\,\left(m_D^{diag.}\right)^2V_d^{L\dagger}\nn
\ea
The current is written
\ba
J_W^{\mu}&=&\bar{u}_L^0\gamma^{\mu}d_L^0\;=\;\bar{u}_LV_u^{L\dagger}\gamma^{\mu}V_d^L\,d_L\nn\\
&\equiv&\bar{u}_L\gamma^{\mu}V_{CKM}d_L
\ea
where the CKM matrix is defined
\ba
V_{CKM}&=&V_u^{L\dagger}\,V_d^L
\ea
Using the Wolfenstein parametrization, the CKM matrix is expressed as follows
\ba
V_{KM}&=&\left(
\begin{array}{ccc}
 1-\frac{\lambda^2}2 & \lambda& A\lambda^3(\rho-\imath\,\eta) \\
 -\lambda  &  1-\frac{\lambda^2}2  & A\lambda^2 \\
A\lambda^3(1-\rho-\imath\,\eta) & -A\lambda^2 & 1
\end{array}
\right)
\ea
with $\lambda\sim 0.2257$ and $A,\rho,\eta$ are order one parameters.

The numerical values of the CKM entries are given by~\cite{Amsler:2008zzb}
\begin{equation}
V_{CKM}=%
\begin{bmatrix}
0.97419 & 0.2257 & 0.00359 \\
-0.2256 & 0.97334 & 0.0415 \\
0.00874 & -0.0407 & 0.999133%
\end{bmatrix}%
~.
\end{equation}%
Taking the logarithm of $V_{CKM}$ we get
\begin{equation}
\ln V_{CKM}=%
\begin{bmatrix}
0 & 0.2277 & -0.012 \\
-0.2277 & 0 & 0.0417 \\
0.0041 & -0.0415 & 0%
\end{bmatrix}%
~.
\end{equation}%
$\ln V_{CKM}$ is not exactly antisymmetric reflecting the fact that $V_{CKM}$ is not
exactly orthogonal because of experimental uncertainties. We may choose
\begin{equation}
\left( \ln V'_{CKM}\right)=%
\begin{bmatrix}
0 & 0.2277 & -0.012 \\
-0.2277 & 0 & 0.0416 \\
0.012 & -0.0416 & 0%
\end{bmatrix}%
\end{equation}%
giving
\begin{equation}
V'_{CKM}=%
\begin{bmatrix}
0.9741 & 0.2259 & -0.0072 \\
-0.2254 & 0.9733 & 0.0426 \\
0.0166 & -0.0399 & 0.9991%
\end{bmatrix}%
\end{equation}%
or
\begin{equation}
\left( \ln V'_{CKM}\right)=%
\begin{bmatrix}
0 & 0.2277 & -0.0041 \\
-0.2277 & 0 & 0.0416 \\
0.0041 & -0.0416 & 0%
\end{bmatrix}%
\end{equation}%
giving
\begin{equation}
V'_{CKM}=%
\begin{bmatrix}
0.9742 & 0.2258 & 0.0007 \\
-0.2256 & 0.9733 & 0.0417 \\
0.0088 & -0.0408 & 0.9991%
\end{bmatrix}%
\end{equation}%
a clearly better choice, suitable for our numerical investigations.
This way the CKM matrix can be written as
\begin{equation}
\ln V_{CKM}=0.0416s_{1}-0.0041s_{2}+0.2277s_{3}~
\end{equation}%
or
\begin{equation}
\ln V_{CKM}=\phi_c\widehat{n}\overrightarrow{s}
\end{equation}%
where $\phi_c=0.231505$ and
\begin{equation}
\widehat{n}=%
\begin{bmatrix}
0.179694 \\
0.0177102 \\
0.983564%
\end{bmatrix}%
\label{vCKM}~\cdot
\end{equation}%
The angle $\phi_c$ and the unit vector $\hat n$ encompass all the information of
the CKM matrix. From a `geometric' perspective, the CKM can be viewed as  rotations
around  three axes defined along the components of $\hat n$. From (\ref{vCKM}) we can
observe that the rotation is predominantly around the third axis, reflect the fact
that the large mixing is between the first two generations.

\subsection{Cholesky form of the  symmetric zero-textures}

The five symmetric texture-zero mass matrices for the up and down quarks
or ref.~\cite{Ramond:1993kv} admit the following Cholesky
form

\begin{enumerate}
\item
\begin{equation}
m_U=%
\begin{bmatrix}
\sqrt{2}\varepsilon ^{6} & 0 & 0 \\
\varepsilon ^{4} & \sqrt{2}\varepsilon ^{6} & 0 \\
0 & 0 & 1%
\end{bmatrix}%
,m_D=%
\begin{bmatrix}
2\varepsilon ^{4} & 0 & 0 \\
2\varepsilon ^{3} & 4\varepsilon ^{3}\sqrt{1+\frac{1}{4}\varepsilon ^{2}} & 0
\\
4\varepsilon ^{3} & \frac{1}{\sqrt{1+\frac{1}{4}\varepsilon ^{2}}} & \frac{%
\varepsilon }{2\sqrt{1+\frac{1}{4}\varepsilon ^{2}}}%
\end{bmatrix}%
\end{equation}

\item
\begin{equation}
m_U=%
\begin{bmatrix}
\varepsilon ^{6} & 0 & 0 \\
0 & \varepsilon ^{2}\sqrt{1+\varepsilon ^{8}} & 0 \\
\varepsilon ^{2} & \frac{1}{\sqrt{1+\varepsilon ^{8}}} & \frac{\varepsilon
^{4}}{\sqrt{1+\varepsilon ^{8}}}%
\end{bmatrix}%
,m_D=%
\begin{bmatrix}
2\varepsilon ^{4} & 0 & 0 \\
2\varepsilon ^{3} & 2\varepsilon ^{3}\sqrt{1+\varepsilon ^{2}} & 0 \\
2\varepsilon ^{3} & \frac{1}{\sqrt{1+\varepsilon ^{2}}} & \frac{\varepsilon
}{\sqrt{1+\varepsilon ^{2}}}%
\end{bmatrix}%
\end{equation}

\item
\begin{equation}
m_U=%
\begin{bmatrix}
\sqrt{2}\varepsilon ^{4} & 0 & 0 \\
0 & \varepsilon ^{4} & 0 \\
1 & 0 & \sqrt{2}\varepsilon ^{4}%
\end{bmatrix}%
m_D=%
\begin{bmatrix}
2\varepsilon ^{4} & 0 & 0 \\
2\varepsilon ^{3} & 4\varepsilon ^{3}\sqrt{1+\frac{1}{4}\varepsilon ^{2}} & 0
\\
4\varepsilon ^{3} & \frac{1}{\sqrt{1+\frac{1}{4}\varepsilon ^{2}}} & \frac{1%
}{2}\frac{\varepsilon }{\sqrt{1+\frac{1}{4}\varepsilon ^{2}}}%
\end{bmatrix}%
\end{equation}

\item
\begin{equation}
m_U=%
\begin{bmatrix}
\sqrt{2}\varepsilon ^{6} & 0 & 0 \\
\sqrt{3}\varepsilon ^{4} & \varepsilon ^{2}\sqrt{1+2\varepsilon ^{8}} & 0 \\
\varepsilon ^{2} & \frac{1}{\sqrt{1+2\varepsilon ^{8}}} & \frac{\sqrt{2}%
\varepsilon ^{4}}{\sqrt{1+2\varepsilon ^{8}}}%
\end{bmatrix}%
,m_D=%
\begin{bmatrix}
2\varepsilon ^{4} & 0 & 0 \\
2\varepsilon ^{3} & 2\varepsilon ^{4} & 0 \\
0 & 0 & 1%
\end{bmatrix}%
\end{equation}

\item
\begin{equation}
m_U=%
\begin{bmatrix}
\varepsilon ^{4} & 0 & 0 \\
\frac{1}{\sqrt{2}}\varepsilon ^{2} & \sqrt{2}\varepsilon ^{4} & 0 \\
1 & \frac{1}{\sqrt{2}}\varepsilon ^{2} & \varepsilon ^{4}%
\end{bmatrix}%
,m_D=%
\begin{bmatrix}
2\varepsilon ^{4} & 0 & 0 \\
2\varepsilon ^{3} & 2\varepsilon ^{4} & 0 \\
0 & 0 & 1%
\end{bmatrix}%
\end{equation}%

\end{enumerate}

\subsection{Textures with two-zeroes }

In this section we will derive the complete list of triangular texture with two-zeroes
with their corresponding orthogonal diagonalizing matrices. As we have seen in the
text, the orthogonal matrix is given by
\begin{equation}
U_{a}=\exp \left[ a\widehat{n}\cdot\overrightarrow{s}\right] =1+\sin a~\widehat{n}%
\cdot\overrightarrow{s}+\left( 1-\cos a\right) \left( \widehat{n}\cdot\overrightarrow{s%
}\right) ^{2}\cdot
\end{equation}%
where $\hat n$ is a unit vector and
$\vec s=(s_1,s_2,s_3)$ with $s_i$ the matrices given in (\ref{sigmas}).
The matrices (\ref{sigmas}) satisfy the algebra
\begin{equation}
\left[ s_{i},~s_{j}\right] =\varepsilon _{ijk}s_{k}
\end{equation}%
and their eigenvalues are $\pm i$ and $0$.
For computational purposes we also notice that
\begin{equation}
\left( \widehat{n}\cdot\overrightarrow{s}\right) ^{3}=-\widehat{n}\cdot\overrightarrow{%
s}
\end{equation}%
and the `commutation' relation
\begin{equation}
\left[ \overrightarrow{a}\cdot\overrightarrow{s},~~\overrightarrow{b}%
\cdot\overrightarrow{s}\right] =\left( \overrightarrow{a}\times \overrightarrow{b}%
\right)\cdot \overrightarrow{s}
\end{equation}%
We have redefined also the vector $\widehat{n}$
\begin{equation}
\widehat{n}\rightarrow \sin \frac{a}{2}\widehat{n}\equiv\overrightarrow{n}~\cdot
\end{equation}%
and the new vector $\overrightarrow{n}$ assumes the components $(a_1,a_2,a_3)$
where now
\begin{equation}
\overrightarrow{n}^{2}=a_{1}^{2}+a_{2}^{2}+a_{3}^{2}=\sin ^{2}\frac{\alpha}{2}%
\cdot
\end{equation}%
The orthogonal matrix $U_{\alpha}$ is finally written
\begin{equation}
U_{\alpha}=1+2\cos \frac{\alpha}{2}\overrightarrow{n}\cdot\overrightarrow{s}+2\left(
\overrightarrow{n}\cdot\overrightarrow{s}\right) ^{2}
\end{equation}%
while in expanded form we get
\begin{equation}
U_{\alpha}=%
\begin{bmatrix}
\cos \alpha+2a_{1}^{2} & 2\left( a_{3}\cos \frac{\alpha}{2}-a_{1}a_{2}\right)  &
2\left( a_{2}\cos \frac{\alpha}{2}+a_{1}a_{3}\right)  \\
-2\left( a_{3}\cos \frac{\alpha}{2}+a_{1}a_{2}\right)  & \cos \alpha+2a_{2}^{2} &
2\left( a_{1}\cos \frac{\alpha}{2}-a_{2}a_{3}\right)  \\
-2\left( a_{2}\cos \frac{\alpha}{2}-a_{1}a_{3}\right)  & -2\left( a_{1}\cos \frac{%
\alpha}{2}+a_{2}a_{3}\right)  & \cos \alpha+2a_{3}^{2}%
\end{bmatrix}%
~.
\end{equation}%
In addition to the cases discussed in the text, zero elements are generated whenever we have
\begin{eqnarray}
a_{1} &=&\pm \cos \frac{\alpha}{2} \\
a_{2} &=&\pm a_{3}=\sqrt{-\frac{1}{2}\cos \alpha}
\end{eqnarray}%
plus permutations. It is obvious that because of the square root
we need $\cos \alpha<0$. The matrix $U_{\alpha}$
acquires four zero elements and one element equal to $\pm 1~$. From the
remaining four elements only two are independent and equal
\begin{eqnarray}
\rho _{1} &=&1+2\cos \alpha \\
\rho _{2} &=&2\sqrt{2}\cos \frac{\alpha}{2}\sqrt{-\cos \alpha}~\cdot
\end{eqnarray}%
Note that since $\cos \alpha <0$, the range of $\rho_1=[-1,1]$ while
\begin{equation}
\rho _{1}^{2}+\rho _{2}^{2}=1
\end{equation}%
i.e. the $2\times 2$ submatrix is orthogonal, thus we may put $\rho_1=\cos\theta, \rho_2=\sin\theta$.

 Below we give a list all the
cases that generate zeroes together with the corresponding $U_{\alpha}$ and $%
MM^{T}$ matrices. To simplify the forthcoming formulae we define the matrix
\ba
{\cal M}&=&\left(\begin{array}{ccc}
m_1^2&0&0\\
0&m_2^2&0\\
0&0&m_3^2\end{array}\right)
\ea
Also,  the sign symbols appearing bellow correspond to the signs of $a_1$ $a_2$ and $a_3$ respectively.
 This way we get ($\cos\theta\ra c, \sin\theta\ra s$):
\begin{itemize}
\item $a_{1}~a_{2}~a_{3}$%
\begin{equation}
\begin{bmatrix}
+ & + & +%
\end{bmatrix},\;%
~%
U=\begin{bmatrix}
c & 0 & s \\
-s & 0 & c \\
0 & -1 & 0%
\end{bmatrix}\nn
\end{equation}%
\begin{equation}
U{\cal M}
U^{T}=%
\begin{bmatrix}
c^{2}m_{1}^{2}+s^{2}m_{3}^{2} & cs\left(
m_{3}^{2}-m_{1}^{2}\right)  & 0 \\
cs\left( m_{3}^{2}-m_{1}^{2}\right)  & c^{2}m_{3}^{2}+s^{2}m_{1}^{2} & 0 \\
0 & 0 & m_{2}^{2}%
\end{bmatrix}\nn
\end{equation}
$\allowbreak $%
\begin{equation}
\begin{bmatrix}
+ & + & -%
\end{bmatrix}\,,%
~%
U=\begin{bmatrix}
c & -s & 0 \\
0 & 0 & 1 \\
-s & -c & 0%
\end{bmatrix}\nn
\end{equation}
\begin{equation}
U
{\cal M}
U^{T}=%
\begin{bmatrix}
c^{2}m_{1}^{2}+s^{2}m_{2}^{2} & 0 & cs\left(
m_{2}^{2}-m_{1}^{2}\right)  \\
0 & m_{3}^{2} & 0 \\
cs\left( m_{2}^{2}-m_{1}^{2}\right)  & 0 & c^{2}m_{2}^{2}+s^{2}m_{1}^{2}%
\end{bmatrix}\nn
\end{equation}
$\allowbreak $%
\begin{equation}
\begin{bmatrix}
+ & - & +%
\end{bmatrix},\,
~%
U=\begin{bmatrix}
c & s & 0 \\
0 & 0 & 1 \\
s & -c & 0%
\end{bmatrix}\nn
\end{equation}%
\begin{equation}
U
{\cal M}
U^{T}=%
\begin{bmatrix}
c^{2}m_{1}^{2}+s^{2}m_{2}^{2} & 0 & s\,c\left(
m_{1}^{2}-m_{2}^{2}\right)  \\
0 & m_{3}^{2} & 0 \\
c\,s _{2}\left( m_{1}^{2}-m_{2}^{2}\right)  & 0 & c^{2}m_{2}^{2}+s^{2}m_{1}^{2}%
\end{bmatrix}\nn
\end{equation}
$\allowbreak $%
\begin{equation}
\begin{bmatrix}
+ & - & -%
\end{bmatrix}\,,%
~%
U=\begin{bmatrix}
c & 0 & -s \\
s & 0 & c\\
0 & -1 & 0%
\end{bmatrix}\nn
\end{equation}%
\begin{equation}
U
{\cal M}
U^{T}=%
\begin{bmatrix}
c^{2}m_{1}^{2}+s^{2}m_{3}^{2} & c\,s\left(
m_{1}^{2}-m_{3}^{2}\right)  & 0 \\
c\,s\left( m_{1}^{2}-m_{3}^{2}\right)  & c^{2}m_{3}^{2}+s^{2}m_{1}^{2} & 0 \\
0 & 0 & m_{2}^{2}%
\end{bmatrix}\nn
\end{equation}
$\allowbreak $%
\begin{equation}
\begin{bmatrix}
- & + & +%
\end{bmatrix}\,,%
~%
U=\begin{bmatrix}
c & s & 0 \\
0 & 0 & -1 \\
-s & c & 0%
\end{bmatrix}
\nn
\end{equation}%
\begin{equation}
U
{\cal M}
U^{T}=%
\begin{bmatrix}
c^{2}m_{1}^{2}+s^{2}m_{2}^{2} & 0 & c\,s\left(
m_{2}^{2}-m_{1}^{2}\right)  \\
0 & m_{3}^{2} & 0 \\
c\,s\left( m_{2}^{2}-m_{1}^{2}\right)  & 0 & c^{2}m_{2}^{2}+s^{2}m_{1}^{2}%
\end{bmatrix}\nn
\end{equation}
$\allowbreak $%
\begin{equation}
\begin{bmatrix}
- & + & -%
\end{bmatrix}\,,
~%
U=\begin{bmatrix}
c & 0 & s \\
s & 0 & -c \\
0 & 1 & 0%
\end{bmatrix}\nn
\end{equation}%
\begin{equation}
~%
U
{\cal M}
U^{T}=%
\begin{bmatrix}
c^{2}m_{1}^{2}+s^{2}m_{3}^{2} & c\,s\left(
m_{1}^{2}-m_{3}^{2}\right)  & 0 \\
c\,s\left( m_{1}^{2}-m_{3}^{2}\right)  & c^{2}m_{3}^{2}+s^{2}m_{1}^{2} & 0 \\
0 & 0 & m_{2}^{2}%
\end{bmatrix}\nn
\end{equation}
$\allowbreak $%
\begin{equation}
\begin{bmatrix}
- & - & +%
\end{bmatrix}\,,%
~%
U=\begin{bmatrix}
c & 0 & -s\\
-s & 0 & -c\\
0 & 1 & 0%
\end{bmatrix}\nn
\end{equation}%
\begin{equation}
~%
U
{\cal M}
~%
U^{T}=%
\begin{bmatrix}
c^{2}m_{1}^{2}+s^{2}m_{3}^{2} & cs\left(
m_{3}^{2}-m_{1}^{2}\right)  & 0 \\
cs\left( m_{3}^{2}-m_{1}^{2}\right)  & c^{2}m_{3}^{2}+s^{2}m_{1}^{2} & 0 \\
0 & 0 & m_{2}^{2}%
\end{bmatrix}\nn
\end{equation}
$\allowbreak $%
\begin{equation}
\begin{bmatrix}
- & - & -%
\end{bmatrix}\,,
~%
U=\begin{bmatrix}
c & -s& 0 \\
0 & 0 & -1 \\
s & c& 0%
\end{bmatrix}\nn
\end{equation}%
\begin{equation}
U
{\cal M}
U^{T}=%
\begin{bmatrix}
c^{2}m_{1}^{2}+s^{2}m_{2}^{2} & 0 & cs\left(
m_{1}^{2}-m_{2}^{2}\right)  \\
0 & m_{3}^{2} & 0 \\
cs\left( m_{1}^{2}-m_{2}^{2}\right)  & 0 & c^{2}m_{2}^{2}+s^{2}m_{1}^{2}%
\end{bmatrix}\nn
\end{equation}
$\allowbreak $

\item $a_{2}~a_{1}~a_{3}$
\begin{equation}
\begin{bmatrix}
+ & + & +%
\end{bmatrix}\,,
~%
U=\begin{bmatrix}
0 & 0 & 1 \\
-s& c & 0 \\
-c & -s & 0%
\end{bmatrix}\nn
\end{equation}%
\begin{equation}
U
{\cal M}
U^{T}=%
\begin{bmatrix}
m_{3}^{2} & 0 & 0 \\
0 & c^{2}m_{2}^{2}+s^{2}m_{1}^{2} & cs\left(
m_{1}^{2}-m_{2}^{2}\right)  \\
0 &cs\left( m_{1}^{2}-m_{2}^{2}\right)  & c^{2}m_{1}^{2}+s^{2}m_{2}^{2}%
\end{bmatrix}\nn
\end{equation}
$\allowbreak $%
\begin{equation}
\begin{bmatrix}
+ & + & -%
\end{bmatrix}\,,
~%
U=\begin{bmatrix}
0 & -s & c \\
0 &c & s \\
-1 & 0 & 0%
\end{bmatrix}\nn
\end{equation}%
\begin{equation}
U
{\cal M}
U^{T}=%
\begin{bmatrix}
c^{2}m_{3}^{2}+s^{2}m_{2}^{2} & cs\left(
m_{3}^{2}-m_{2}^{2}\right)  & 0 \\
cs\left( m_{3}^{2}-m_{2}^{2}\right)  & c^{2}m_{2}^{2}+s^{2}m_{3}^{2} & 0 \\
0 & 0 & m_{1}^{2}%
\end{bmatrix}\nn
\end{equation}
$\allowbreak $%
\begin{equation}
\begin{bmatrix}
+ & - & +%
\end{bmatrix}\,,
~%
U=\begin{bmatrix}
0 &s & c \\
0 & c & -s \\
-1 & 0 & 0%
\end{bmatrix}\nn
\end{equation}%
\begin{equation}
U
{\cal M} U^{T}=%
\begin{bmatrix}
c^{2}m_{3}^{2}+s^{2}m_{2}^{2} & cs\left(
m_{2}^{2}-m_{3}^{2}\right)  & 0 \\
cs\left( m_{2}^{2}-m_{3}^{2}\right)  & c^{2}m_{2}^{2}+s^{2}m_{3}^{2} & 0 \\
0 & 0 & m_{1}^{2}%
\end{bmatrix}\nn
\end{equation}
$\allowbreak $%
\begin{equation}
\begin{bmatrix}
+ & - & -%
\end{bmatrix}\,,
~%
U=\begin{bmatrix}
0 & 0 & 1 \\
s& c & 0 \\
-c & s& 0%
\end{bmatrix}\nn
\end{equation}%
\begin{equation}
U{\cal M}U^{T}=%
\begin{bmatrix}
m_{3}^{2} & 0 & 0 \\
0 & c^{2}m_{2}^{2}+s^{2}m_{1}^{2} & c\,s\left(
m_{2}^{2}-m_{1}^{2}\right)  \\
0 & cs\left( m_{2}^{2}-m_{1}^{2}\right)  & c^{2}m_{1}^{2}+s^{2}m_{2}^{2}%
\end{bmatrix}\nn
\end{equation}
$\allowbreak $%
\begin{equation}
\begin{bmatrix}
- & + & +%
\end{bmatrix}\,,
~%
U=\begin{bmatrix}
0 & s & -c\\
0 & c & s \\
1 & 0 & 0%
\end{bmatrix}\nn
\end{equation}%
\begin{equation}
U{\cal M}U^{T}=%
\begin{bmatrix}
c^{2}m_{3}^{2}+s^{2}m_{2}^{2} & cs\left(
m_{2}^{2}-m_{3}^{2}\right)  & 0 \\
cs\left( m_{2}^{2}-m_{3}^{2}\right)  & c^{2}m_{2}^{2}+s^{2}m_{3}^{2} & 0 \\
0 & 0 & m_{1}^{2}%
\end{bmatrix}\nn
\end{equation}
$\allowbreak $%
\begin{equation}
\begin{bmatrix}
- & + & -%
\end{bmatrix}\,,
~%
U=\begin{bmatrix}
0 & 0 & -1 \\
s &c & 0 \\
c & -s & 0%
\end{bmatrix}\nn
\end{equation}%
\begin{equation}
U{\cal M}U^{T}=%
\begin{bmatrix}
m_{3}^{2} & 0 & 0 \\
0 & c^{2}m_{2}^{2}+s^{2}m_{1}^{2} & cs\left(
m_{1}^{2}-m_{2}^{2}\right)  \\
0 & cs\left( m_{1}^{2}-m_{2}^{2}\right)  & c^{2}m_{1}^{2}+s^{2}m_{2}^{2}%
\end{bmatrix}\nn
\end{equation}
$\allowbreak $%
\begin{equation}
\begin{bmatrix}
- & - & +%
\end{bmatrix}\,,
~%
U=\begin{bmatrix}
0 & 0 & -1 \\
-s & c & 0 \\
c & s& 0%
\end{bmatrix}\nn
\end{equation}%
\begin{equation}
U{\cal M}U^{T}=%
\begin{bmatrix}
m_{3}^{2} & 0 & 0 \\
0 & c^{2}m_{2}^{2}+s^{2}m_{1}^{2} & cs\left(
m_{2}^{2}-m_{1}^{2}\right)  \\
0 & cs\left( m_{2}^{2}-m_{1}^{2}\right)  & c^{2}m_{1}^{2}+s{2}m_{2}^{2}%
\end{bmatrix}\nn
\end{equation}
$\allowbreak $%
\begin{equation}
\begin{bmatrix}
- & - & -%
\end{bmatrix}\,,
~%
U=\begin{bmatrix}
0 & -s & -c\\
0 & c & -s \\
1 & 0 & 0%
\end{bmatrix}\nn
\end{equation}%
\begin{equation}
U{\cal M}U^{T}=%
\begin{bmatrix}
c^{2}m_{3}^{2}+s^{2}m_{2}^{2} & cs\left(
m_{3}^{2}-m_{2}^{2}\right)  & 0 \\
cs\left( m_{3}^{2}-m_{2}^{2}\right)  & c^{2}m_{2}^{2}+s^{2}m_{3}^{2} & 0 \\
0 & 0 & m_{1}^{2}%
\end{bmatrix}\nn
\end{equation}
$\allowbreak $

\item $a_{3}~a_{1}~a_{2}$%
\begin{equation}
\begin{bmatrix}
+ & + & +%
\end{bmatrix}\,,
~%
U=\begin{bmatrix}
0 & c & s \\
-1 & 0 & 0 \\
0 & -s & c%
\end{bmatrix}\nn
\end{equation}%
\begin{equation}
U{\cal M}U^{T}=%
\begin{bmatrix}
c^{2}m_{2}^{2}+s^{2}m_{3}^{2} & 0 & cs\left(
m_{3}^{2}-m_{2}^{2}\right)  \\
0 & m_{1}^{2} & 0 \\
cs\left( m_{3}^{2}-m_{2}^{2}\right)  & 0 & c^{2}m_{3}^{2}+s^{2}m_{2}^{2}%
\end{bmatrix}\nn
\end{equation}
$\allowbreak $%
\begin{equation}
\begin{bmatrix}
+ & + & -%
\end{bmatrix}\,,
~%
U=\begin{bmatrix}
0 & 1 & 0 \\
-c& 0 & s \\
s & 0 & c%
\end{bmatrix}\nn
\end{equation}%
\begin{equation}
U{\cal M}
U^{T}=%
\begin{bmatrix}
m_{2}^{2} & 0 & 0 \\
0 & c^{2}m_{1}^{2}+s^{2}m_{3}^{2} & cs\left(
m_{3}^{2}-m_{1}^{2}\right)  \\
0 & cs\left( m_{3}^{2}-m_{1}^{2}\right)  & c^{2}m_{3}^{2}+s^{2}m_{1}^{2}%
\end{bmatrix}\nn
\end{equation}
$\allowbreak $%
\begin{equation}
\begin{bmatrix}
+ & - & +%
\end{bmatrix}\,,
~%
U=\begin{bmatrix}
0 & 1 & 0 \\
-c & 0 & -s \\
-s & 0 &c%
\end{bmatrix}\nn
\end{equation}%
\begin{equation}
~%
U{\cal M}U^{T}=%
\begin{bmatrix}
m_{2}^{2} & 0 & 0 \\
0 & c^{2}m_{1}^{2}+s^{2}m_{3}^{2} &cs\left(
m_{1}^{2}-m_{3}^{2}\right)  \\
0 & cs\left( m_{1}^{2}-m_{3}^{2}\right)  & c^{2}m_{3}^{2}+s^{2}m_{1}^{2}%
\end{bmatrix}\nn
\end{equation}
$\allowbreak $%
\begin{equation}
\begin{bmatrix}
+ & - & -%
\end{bmatrix}\,,
~%
U=\begin{bmatrix}
0 &c& -s \\
-1 & 0 & 0 \\
0 & s&c%
\end{bmatrix}\nn
\end{equation}%
\begin{equation}
U{\cal M}U^{T}=%
\begin{bmatrix}
c^{2}m_{2}^{2}+s^{2}m_{3}^{2} & 0 &cs\left(
m_{2}^{2}-m_{3}^{2}\right)  \\
0 & m_{1}^{2} & 0 \\
cs\left( m_{2}^{2}-m_{3}^{2}\right)  & 0 & c^{2}m_{3}^{2}+s^{2}m_{2}^{2}%
\end{bmatrix}\nn
\end{equation}
$\allowbreak $%
\begin{equation}
\begin{bmatrix}
- & + & +%
\end{bmatrix}\,,
~%
U=\begin{bmatrix}
0 & -1 & 0 \\
c& 0 & s\\
-s& 0 & c%
\end{bmatrix}\nn
\end{equation}%
\begin{equation}
U{\cal M}U^{T}=%
\begin{bmatrix}
m_{2}^{2} & 0 & 0 \\
0 & c^{2}m_{1}^{2}+s^{2}m_{3}^{2} & cs\left(
m_{3}^{2}-m_{1}^{2}\right)  \\
0 & cs\left( m_{3}^{2}-m_{1}^{2}\right)  & c^{2}m_{3}^{2}+s^{2}m_{1}^{2}%
\end{bmatrix}\nn
\end{equation}
$\allowbreak $%
\begin{equation}
\begin{bmatrix}
- & + & -%
\end{bmatrix}\,,
~%
U=\begin{bmatrix}
0 & -c & -s\\
1 & 0 & 0 \\
0 & -s & c%
\end{bmatrix}\nn
\end{equation}%
\begin{equation}
U{\cal M}U^{T}=%
\begin{bmatrix}
c^{2}m_{2}^{2}+s^{2}m_{3}^{2} & 0 & cs\left(
m_{2}^{2}-m_{3}^{2}\right)  \\
0 & m_{1}^{2} & 0 \\
cs\left( m_{2}^{2}-m_{3}^{2}\right)  & 0 &c^{2}m_{3}^{2}+s^{2}m_{2}^{2}%
\end{bmatrix}\nn
\end{equation}
$\allowbreak $%
\begin{equation}
\begin{bmatrix}
- & - & +%
\end{bmatrix}\,,
~%
U=\begin{bmatrix}
0 & -c& s\\
1 & 0 & 0 \\
0 & s &c%
\end{bmatrix}\nn
\end{equation}%
\begin{equation}
U{\cal M}U^{T}=%
\begin{bmatrix}
c^{2}m_{2}^{2}+s^{2}m_{3}^{2} & 0 & cs\left(
m_{3}^{2}-m_{2}^{2}\right)  \\
0 & m_{1}^{2} & 0 \\
cs\left( m_{3}^{2}-m_{2}^{2}\right)  & 0 &c^{2}m_{3}^{2}+s^{2}m_{2}^{2}%
\end{bmatrix}\nn
\end{equation}
$\allowbreak $%
\begin{equation}
\begin{bmatrix}
- & - & -%
\end{bmatrix}\,,
~%
U=\begin{bmatrix}
0 & -1 & 0 \\
c& 0 & -s\\
s & 0 & c%
\end{bmatrix}\nn
\end{equation}%
\begin{equation}
U{\cal M}U^{T}=%
\begin{bmatrix}
m_{2}^{2} & 0 & 0 \\
0 & c^{2}m_{1}^{2}+s^{2}m_{3}^{2} & cs\left(
m_{1}^{2}-m_{3}^{2}\right)  \\
0 & cs\left( m_{1}^{2}-m_{3}^{2}\right)  &c^{2}m_{3}^{2}+s^{2}m_{1}^{2}%
\end{bmatrix}\nn
\end{equation}
$\allowbreak $
\end{itemize}

\subsection{One zero textures.}

If we want the triangular matrix to have one zero element we must address
the following geometrical problem. Given an orthonormal basis in the 3-d
space $\overrightarrow{a}$,$\overrightarrow{b}$ and $\overrightarrow{c}=%
\overrightarrow{a}\times \overrightarrow{b}$ $\ $and a matrix $M$ (diagonal
in our case) we must satisfy the condition%
\begin{equation}
\overrightarrow{a}M\overrightarrow{b}=0\label{1ztc}\cdot
\end{equation}%
If this relation is to hold true we must have%
\begin{equation}
M\overrightarrow{b}=\left( \overrightarrow{b}M\overrightarrow{b}\right)
\overrightarrow{b}+\left( \overrightarrow{c}M\overrightarrow{b}\right)
\overrightarrow{c}
\end{equation}%
or
\begin{equation}
\left[ M-\overrightarrow{b}M\overrightarrow{b}\right] \overrightarrow{b}=%
\left( \overrightarrow{c}M\overrightarrow{b}\right)\overrightarrow{c}~\cdot\label{140}
\end{equation}%
Furthermore, in order to check whether the vector $\overrightarrow{c}$
admits also zero components  we must investigate the relation
\begin{equation}
M=\overrightarrow{b}M\overrightarrow{b}~\cdot
\end{equation}%
This relation in components gives
\begin{eqnarray}
b_{2}^{2}\left( m_{1}^{2}-m_{2}^{2}\right) &=&b_{3}^{2}\left(
m_{3}^{2}-m_{1}^{2}\right) \\
b_{1}^{2}\left( m_{2}^{2}-m_{1}^{2}\right) &=&b_{3}^{2}\left(
m_{3}^{2}-m_{2}^{2}\right)\label{vecc2} \\
b_{1}^{2}\left( m_{3}^{2}-m_{1}^{2}\right) &=&b_{2}^{2}\left(
m_{2}^{2}-m_{3}^{2}\right) ~\cdot
\end{eqnarray}%
Taking into account the charged fermion mass hierarchy $m_3>m_2>m_1$,
 we observe that only the second line (\ref{vecc2}) can be satisfied so
 that only $c_{2}$ can vanish.%

 Returning to (\ref{140}) and taking the ratios of its components
  we obtain
\begin{equation}
\frac{c_{1}}{c_{2}}=\frac{m_{1}^{2}-\overrightarrow{b}M\overrightarrow{b}}{%
m_{2}^{2}-\overrightarrow{b}M\overrightarrow{b}}\frac{b_{1}}{b_{2}}\,\equiv\,a_{12}\frac{b_{1}}{b_{2}}
\end{equation}%
\begin{equation}
\frac{c_{2}}{c_{3}}=\frac{m_{2}^{2}-\overrightarrow{b}M\overrightarrow{b}}{%
m_{3}^{2}-\overrightarrow{b}M\overrightarrow{b}}\frac{b_{2}}{b_{3}}\,\equiv\,a_{23}\frac{b_{2}}{b_{3}}
\end{equation}%
\begin{equation}
\frac{c_{1}}{c_{3}}=\frac{m_{1}^{2}-\overrightarrow{b}M\overrightarrow{b}}{%
m_{3}^{2}-\overrightarrow{b}M\overrightarrow{b}}\frac{b_{1}}{b_{3}}\,\equiv\,a_{13}\frac{b_{1}}{b_{3}}~\cdot
\end{equation}%
in a self-explanatory notation. Solving for $c_{1},c_{2}$ we get
\begin{eqnarray}
c_{2} &=&a_{23}\frac{b_{2}}{b_{3}}c_{3} \\
c_{1} &=&a_{13}\frac{b_{1}}{b_{3}}c_{3}
\end{eqnarray}%
We observe that the orthogonality condition
\begin{equation}
b_{1}c_{1}+b_{2}c_{2}+b_{3}c_{3}=0
\end{equation}%
is satisfied automatically provided that
\begin{equation}
b_{1}^{2}+b_{2}^{2}+b_{3}^{2}=1~\cdot
\end{equation}%
Also, we find that
\begin{equation}
c_{1}^{2}+c_{2}^{2}+c_{3}^{2}=c_{3}^{2}\left[ 1+\frac{%
a_{23}^{2}b_{2}^{2}+a_{13}^{2}b_{1}^{2}}{b_{3}^{2}}\right] =1
\end{equation}%
so that the $c_3$ component can be expressed as a function of  the mass eigenstates
and the vector $\overrightarrow{b}$ as follows
\begin{equation}
c_{3}=\pm \frac{b_{3}}{\sqrt{%
a_{13}^{2}b_{1}^{2}+a_{23}^{2}b_{2}^{2}+b_{3}^{2}}}~\cdot
\end{equation}%
The remaining two components $c_{1,2}$ follow immediately, thus we finally get:
\begin{equation}
c_{1}=a_{13}\frac{b_{1}}{\sqrt{%
a_{13}^{2}b_{1}^{2}+a_{23}^{2}b_{2}^{2}+b_{3}^{2}}}
\end{equation}%
\begin{equation}
c_{2}=a_{23}\frac{b_{2}}{\sqrt{%
a_{13}^{2}b_{1}^{2}+a_{23}^{2}b_{2}^{2}+b_{3}^{2}}}~
\end{equation}%
\begin{equation}
c_{3}=\pm \frac{b_{3}}{\sqrt{%
a_{13}^{2}b_{1}^{2}+a_{23}^{2}b_{2}^{2}+b_{3}^{2}}}~\cdot
\end{equation}%
Depending on the sign of $c_3$, the vector $\overrightarrow{a}=\overrightarrow{b}\times \overrightarrow{c}$
equals
\begin{eqnarray}
a_{1} &=&\frac{\left( 1-a_{23}\right) b_{2}b_{3}}{%
\sqrt{a_{13}^{2}b_{1}^{2}+a_{23}^{2}b_{2}^{2}+b_{3}^{2}}} \\
a_{2} &=&\frac{\left( a_{13}-1\right) b_{1}b_{3}}{%
\sqrt{a_{13}^{2}b_{1}^{2}+a_{23}^{2}b_{2}^{2}+b_{3}^{2}}} \\
a_{3} &=&\frac{\left( a_{23}-a_{13}\right) b_{1}b_{2}}{%
\sqrt{a_{13}^{2}b_{1}^{2}+a_{23}^{2}2b_{2}^{2}+b_{3}^{2}}}
\end{eqnarray}%
or
\begin{eqnarray}
a_{1} &=&-\frac{\left( 1+a_{23}\right) b_{2}b_{3}}{%
\sqrt{a_{13}^{2}b_{1}^{2}+a_{23}^{2}b_{2}^{2}+b_{3}^{2}}} \\
a_{2} &=&\frac{\left( 1+a_{13}\right) b_{1}b_{3}}{%
\sqrt{a_{13}^{2}b_{1}^{2}+a_{23}^{2}b_{2}^{2}+b_{3}^{2}}} \\
a_{3} &=&\frac{\left( a_{23}-a_{13}\right) b_{1}b_{2}}{%
\sqrt{a_{13}^{2}b_{1}^{2}+a_{23}^{2}b_{2}^{2}+b_{3}^{2}}}\cdot
\end{eqnarray}%
Thus the formulae  for the $a_i,c_i$ components constitute the general solution to the
one-zero texture expressed by the condition (\ref{1ztc}).

Going back to the definitions of the triangular (Cholesky) matrix (\ref{Chol}),
 we distinguish the following three cases with respect to the three off-diagonal entries:

$i$) To apply the above formulae for a zero $\{21\}$ element in the triangular matrix
\begin{equation}
\overrightarrow{\xi}_{2}\cdot \overrightarrow{e}_{1}=\frac{%
u_{11}u_{21}m_{1}^{2}+u_{22}u_{12}m_{2}^{2}+u_{13}u_{23}m_{3}^{2}}{\sqrt{%
u_{11}^{2}m_{1}^{2}+u_{12}^{2}m_{2}^{2}+u_{13}^{2}m_{3}^{2}}}
\end{equation}%
we make the identifications
\begin{eqnarray}
\overrightarrow{a} &=&\left[ u_{11},u_{12},u_{13}\right] \\
\overrightarrow{b} &=&\left[ u_{21},u_{22},u_{23}\right] \\
\overrightarrow{c} &=&\left[ u_{31},u_{32},u_{33}\right] \cdot
\end{eqnarray}%
At the same time,  the only allowed zero element in the orthogonal matrix
is  $u_{32}$, in accordance with (\ref{vecc2}).

$ii$) For a zero $\{31\}$ element
\begin{equation}
\overrightarrow{\xi}_{3}\cdot \overrightarrow{e}_{1}=\frac{%
u_{11}u_{31}m_{1}^{2}+u_{32}u_{12}m_{2}^{2}+u_{33}u_{13}m_{3}^{2}}{\sqrt{%
u_{11}^{2}m_{1}^{2}+u_{12}^{2}m_{2}^{2}+u_{13}^{2}m_{3}^{2}}}
\end{equation}%
we make the following substitutions in our general results
\begin{eqnarray}
\overrightarrow{a} &=&\left[ u_{11},u_{12},u_{13}\right] \\
\overrightarrow{b} &=&\left[ u_{31},u_{32},u_{33}\right] \\
\overrightarrow{c} &=&-\left[ u_{21},u_{22},u_{23}\right] \cdot
\end{eqnarray}%
The zero element of the orthogonal matrix in this case  according
to (\ref{vecc2}) is $u_{22}=0$.%

$iii$) Finally, the application for the $\{32\}$ element
\begin{equation}
\overrightarrow{\xi}_{3}\cdot \overrightarrow{e}_{2}=-\frac{%
u_{23}u_{33}m_{1}^{2}m_{2}^{2}+u_{21}u_{31}m_{2}^{2}m_{3}^{2}+u_{22}u_{32}m_{1}^{2}m_{3}^{2}%
}{\sqrt{u_{11}^{2}m_{1}^{2}+u_{12}^{2}m_{2}^{2}+u_{13}^{2}m_{3}^{2}}\sqrt{%
u_{33}^{2}m_{1}^{2}m_{2}^{2}+u_{31}^{2}m_{2}^{2}m_{3}^{2}+u_{32}^{2}m_{1}^{2}m_{3}^{2}%
}}~.
\end{equation}%
implies that
\begin{eqnarray}
\overrightarrow{a} &=&\left[ u_{21},u_{22},u_{23}\right] \\
\overrightarrow{b} &=&\left[ u_{31},u_{32},u_{33}\right] \\
\overrightarrow{c} &=&\left[ u_{11},u_{12},u_{13}\right] \cdot
\end{eqnarray}%
while for this particular case we also have to substitute
\begin{equation}
m_{1}^{2}\rightarrow m_{2}^{2}m_{3}^{2}
\end{equation}%
\begin{equation}
m_{2}^{2}\rightarrow m_{1}^{2}m_{3}^{2}
\end{equation}%
\begin{equation}
m_{3}^{2}\rightarrow m_{1}^{2}m_{2}^{2}\cdot
\end{equation}%
The only possible  zero element of the orthogonal matrix in this case is $u_{12}$.


\newpage
\begin{thebibliography}{99}


\bibitem{Antonelli:2009ws}
  M.~Antonelli {\it et al.},
  ``Flavor Physics in the Quark Sector,''
  arXiv:0907.5386 [hep-ph].



\bibitem{Froggatt:1978nt}
  C.~D.~Froggatt and H.~B.~Nielsen,
  ``Hierarchy Of Quark Masses, Cabibbo Angles And CP Violation,''
  Nucl.\ Phys.\  B {\bf 147} (1979) 277.
  \\
  D.~B.~Kaplan and M.~Schmaltz,
  ``Flavor unification and discrete nonAbelian symmetries,''
  Phys.\ Rev.\  D {\bf 49} (1994) 3741
  [arXiv:hep-ph/9311281].
  \\
  L.~E.~Ibanez and G.~G.~Ross,
  ``Fermion masses and mixing angles from gauge symmetries,''
  Phys.\ Lett.\  B {\bf 332} (1994) 100
  [arXiv:hep-ph/9403338].
  \\
  P.~Binetruy and P.~Ramond,
  ``Yukawa textures and anomalies,''
  Phys.\ Lett.\  B {\bf 350} (1995) 49
  [arXiv:hep-ph/9412385].
  \\
  E.~Dudas, S.~Pokorski and C.~A.~Savoy,
  ``Yukawa matrices from a spontaneously broken Abelian symmetry,''
  Phys.\ Lett.\  B {\bf 356} (1995) 45
  [arXiv:hep-ph/9504292].
  \\
  P.~Binetruy, S.~Lavignac and P.~Ramond,
  ``Yukawa textures with an anomalous horizontal Abelian symmetry,''
  Nucl.\ Phys.\  B {\bf 477} (1996) 353
  [arXiv:hep-ph/9601243].
  \\
  G.~K.~Leontaris and J.~Rizos,
  ``New fermion mass textures from anomalous U(1) symmetries with baryon  and
  lepton number conservation,''
  Nucl.\ Phys.\  B {\bf 567} (2000) 32
  [arXiv:hep-ph/9909206].
  \\
  S.~F.~King and G.~G.~Ross,
  ``Fermion masses and mixing angles from SU(3) family symmetry,''
  Phys.\ Lett.\  B {\bf 520} (2001) 243
  [arXiv:hep-ph/0108112].


\bibitem{Antoniadis:1989zy}
  I.~Antoniadis, J.~R.~Ellis, J.~S.~Hagelin and D.~V.~Nanopoulos,
  ``The Flipped $SU(5) \times U(1)$ String Model Revamped,''
  Phys.\ Lett.\  B {\bf 231} (1989) 65.

\bibitem{Antoniadis:1990hb}
  I.~Antoniadis, G.~K.~Leontaris and J.~Rizos,
  ``A Three generation $SU(4) \times O(4)$ string model,''
  Phys.\ Lett.\  B {\bf 245} (1990) 161.
  \\
  G.~K.~Leontaris and J.~Rizos,
  ``N=1 supersymmetric $SU(4)\times SU(2)_L\times SU(2)_R$ effective theory from the weakly
  coupled heterotic superstring,''
  Nucl.\ Phys.\  B {\bf 554} (1999) 3
  [arXiv:hep-th/9901098].

\bibitem{Faraggi:1991jr}
  A.~E.~Faraggi,
  ``A New standard - like model in the four-dimensional free fermionic string
  formulation,''
  Phys.\ Lett.\  B {\bf 278} (1992) 131.

  \bibitem{Leontaris:1999ce}
  G.~K.~Leontaris and J.~Rizos,
  ``N=1 supersymmetric SU(4)xSU(2)LxSU(2)R effective theory from the weakly
  coupled heterotic superstring,''
  Nucl.\ Phys.\  B {\bf 554} (1999) 3
  [arXiv:hep-th/9901098].


\bibitem{Faraggi:2006bc}
  A.~E.~Faraggi, C.~Kounnas and J.~Rizos,
  ``Chiral family classification of fermionic Z(2) x Z(2) heterotic orbifold
  models,''
  Phys.\ Lett.\  B {\bf 648} (2007) 84
  [arXiv:hep-th/0606144].


\bibitem{Ramond:1993kv}
  P.~Ramond, R.~G.~Roberts and G.~G.~Ross,
  ``Stitching the Yukawa quilt,''
  Nucl.\ Phys.\  B {\bf 406} (1993) 19
  [arXiv:hep-ph/9303320].

\bibitem{Ibanez:2008my}
  L.~E.~Ibanez and R.~Richter,
  ``Stringy Instantons and Yukawa Couplings in MSSM-like Orientifold Models,''
  JHEP {\bf 0903} (2009) 090
  [arXiv:0811.1583 [hep-th]].


  \bibitem{Leontaris:2009ci}
  G.~K.~Leontaris,
  ``Instanton induced charged fermion and neutrino masses in a minimal Standard
  Model scenario from intersecting D-branes,''
  arXiv:0903.3691 [hep-ph].


\bibitem{Anastasopoulos:2009mr}
  P.~Anastasopoulos, E.~Kiritsis and A.~Lionetto,
  ``On mass hierarchies in orientifold vacua,''
  arXiv:0905.3044 [hep-th].


  \bibitem{Cvetic:2009yh}
  M.~Cvetic, J.~Halverson and R.~Richter,
  ``Realistic Yukawa structures from orientifold compactifications,''
  arXiv:0905.3379 [hep-th].
  
  \bibitem{Kokorelis:2008ce}
  C.~Kokorelis,
  ``On the (Non) Perturbative Origin of Quark Masses in D-brane GUT Models,''
  arXiv:0812.4804 [hep-th].


    %
\bibitem{Kiritsis:2009sf}
  E.~Kiritsis, M.~Lennek and B.~Schellekens,
  ``SU(5) orientifolds, Yukawa couplings, Stringy Instantons and Proton
  Decay,''
  arXiv:0909.0271 [hep-th].


\bibitem{Cvetic:2009ez}
  M.~Cvetic, J.~Halverson and R.~Richter,
  ``Mass Hierarchies from MSSM Orientifold Compactifications,''
  arXiv:0909.4292 [hep-th].

  \bibitem{Beasley:2008dc}
  C.~Beasley, J.~J.~Heckman and C.~Vafa,
  ``GUTs and Exceptional Branes in F-theory - I,''
  JHEP {\bf 0901} (2009) 058
  [arXiv:0802.3391 [hep-th]].

  %
\bibitem{Beasley:2008kw}
  C.~Beasley, J.~J.~Heckman and C.~Vafa,
  ``GUTs and Exceptional Branes in F-theory - II: Experimental Predictions,''
  JHEP {\bf 0901} (2009) 059
  [arXiv:0806.0102 [hep-th]].

%
\bibitem{Heckman:2008qa}
  J.~J.~Heckman and C.~Vafa,
  ``Flavor Hierarchy From F-theory,''
  arXiv:0811.2417 [hep-th].

\bibitem{Font:2008id}
  A.~Font and L.~E.~Ibanez,
  ``Yukawa Structure from U(1) Fluxes in F-theory Grand Unification,''
  JHEP {\bf 0902} (2009) 016
  [arXiv:0811.2157 [hep-th]].

\bibitem{Randall:2009dw}
  L.~Randall and D.~Simmons-Duffin,
  ``Quark and Lepton Flavor Physics from F-Theory,''
  arXiv:0904.1584 [hep-ph].


\bibitem{Ibanez:2001nd}
  L.~E.~Ibanez, F.~Marchesano and R.~Rabadan,
  ``Getting just the standard model at intersecting branes,''
  JHEP {\bf 0111} (2001) 002
  [arXiv:hep-th/0105155].
  \\
  R.~Blumenhagen, B.~Kors, D.~Lust and T.~Ott,
  ``The standard model from stable intersecting brane world orbifolds,''
  Nucl.\ Phys.\  B {\bf 616} (2001) 3
  [arXiv:hep-th/0107138].
  \\
  R.~Blumenhagen, M.~Cvetic, P.~Langacker and G.~Shiu,
  ``Toward realistic intersecting D-brane models,''
  Ann.\ Rev.\ Nucl.\ Part.\ Sci.\  {\bf 55} (2005) 71
  [arXiv:hep-th/0502005].
  

    \bibitem{Blumenhagen:2006xt}
  R.~Blumenhagen, M.~Cvetic and T.~Weigand,
  ``Spacetime instanton corrections in 4D string vacua - the seesaw mechanism
  for D-brane models,''
  Nucl.\ Phys.\  B {\bf 771} (2007) 113
  [arXiv:hep-th/0609191].


  \bibitem{Ibanez:2006da}
  L.~E.~Ibanez and A.~M.~Uranga,
  ``Neutrino Majorana masses from string theory instanton effects,''
  JHEP {\bf 0703} (2007) 052
  [arXiv:hep-th/0609213].

  \bibitem{Wolfenstein:1983yz}
  L.~Wolfenstein,
  ``Parametrization Of The Kobayashi-Maskawa Matrix,''
  Phys.\ Rev.\ Lett.\  {\bf 51} (1983) 1945.


\bibitem{Gantmacher}
F.R. ~Gantmacher, ``The Theory of Matrices'', Chealsea P.C., New York 1959.

  %
\bibitem{Amsler:2008zzb}
  C.~Amsler {\it et al.}  [Particle Data Group],
  Phys.\ Lett.\  B {\bf 667} (2008) 1.



  \end{thebibliography}
\end{document}